# Ferroelectric nematic liquid crystalline phases

Nerea Sebastián[1], and Martin Čopič[1,2], Alenka Mertelj[1]
[1]*J. Stefan Institute, SI-1000 Ljubljana, Slovenia*
[2]*University of Ljubljana, Faculty of Mathematics and Physics, Ljubljana, Slovenia*

(Date: 2022-07-27)

Recent experimental realization of ferroelectric nematic liquid crystalline phases stimulated material development and numerous experimental studies of these new phases, guided by their fundamental and applicative interest. In this Perspective, we give an overview of this emerging field by linking history and theoretical predictions to a general outlook of the development and properties of the materials exhibiting ferroelectric nematic phases. We will highlight the most relevant observations up-to-date, e.g., giant dielectric permittivity values, polarization values an order of magnitude larger than in classical ferroelectric liquid crystals, and nonlinear optical coefficients comparable to several ferroelectric solid materials. Key observations of anchoring and electro-optic behavior will also be examined. The collected contributions lead to a final discussion on open challenges in materials development, theoretical description, experimental explorations, and possible applications of the ferroelectric phases.

**Subject Areas:** Soft Matter, Materials Science

## I. INTRODUCTION

The term liquid crystals (LCs) refers to a state of matter, intermediate between isotropic liquid and solid state, whose properties fall in between. The liquid crystalline state is not unique, and there is a rich variety of phases, of which a given material can exhibit one or several. The least ordered of such phases is the nematic phase, in which the constituents exhibit long-range orientational order but no long-range translational order. The simple uniaxial nematic phase (N) formed by nonchiral molecules, is the most widely studied and technologically exploited of the liquid crystalline phases. On the other hand, chiral molecules are known to form a twisted (chiral) nematic phase (N*) and three so called "blue" phases characterized by a lattice of line defects. It has not been until the last decade that new nematic phases have been added to this exclusive list. In 2011 the twist-bend nematic phase ($N_{TB}$), predicted years before [1,2], was discovered [3–5] for achiral flexible dimer molecules. And more recently, thermotropic polar nematic phases have been experimentally realized, which are the subject of this paper.

The idea that a nematic phase could be ferroelectric was first suggested by Born in 1916 [6]. In his paper, Born proposed that experimentally observed phase transition from an isotropic to a nematic phase is a transition from a paraelectric to a ferroelectric phase and that such phase transition can be described by a mean field theory analogous to the Langevin-Weiss theory for the phase transition from a paramagnetic to a ferromagnetic state. Decades later, Frank [7] discussed the consequence of polar order on elastic free energy density and realized that in the case that the constituents of the nematic phase lack head-tail symmetry, polar order and splay deformation are connected. This was later in more detail studied by Plainer and Brand [8], who showed that the phase exhibiting spontaneous splay deformation could be the thermodynamically stable ground state for some temperature interval. Palffy-Muhoray *et al* [9] investigated the realizability of the ferroelectric nematic phase in terms of molecular constraints and concluded that with careful molecular design, disk-shaped molecules can form a uniform nematic phase which is also ferroelectric. Using Monte Carlo simulations, Berardi *et al* [10,11] showed that axially symmetric tapered, conical constituents with asymmetric interaction of the two ends can form polar phases even if they do not carry a dipole moment. Adding a small axial dipole results in a ferroelectric nematic phase. With increasing dipole moment, polar order starts to disappear, while simultaneously complex domain structures with local polar order arise [11].

Besides all the theoretical analysis, experimentally, the ferroelectric nematic phase remained elusive for a long time. Polar order has been observed in some rigid linear polymers when they were sufficiently polymerized [12,13] and in a lyotropic cholesteric phase of poly L-glutamate [14]. At last in the year 2017, 101 years after Born paper, two types of low molecular mass materials exhibiting ferroelectric phases were reported. Nishikawa *et al*. discovered that a material having a 1,3-dioxane unit in the mesogenic core, which they named DIO, exhibits three nematic phases [15]. They showed that the lowest temperature phase exhibits polar arrangement of molecules, which results in large spontaneous electric polarization along the director [15]. Mandle *et al*. designed a family of rod-like materials with a terminal nitro group and alkyloxy side group, which exhibit two nematic phases [16,17]. The representative material of this family, RM734, was later studied in the detail. First, it was shown by A. Mertelj et al. that at the phase transition between the nematic phases, simultaneously strong softening of splay deformation and growing ferroelectric order occur, resulting in a ferroelectric nematic order on the macroscopic scale [18,19]. X. Chen et al. then showed that the low-temperature phase has large ferroelectric domains with saturated polarization of the same order of magnitude as the material DIO [20].

In this Perspective, we aim to give an overview of the recent experimental advances in the field of ferroelectric nematic liquid crystals and frame them in the existing theoretical background. The manuscript is organized as follows. First, a brief description of nematic phases and comprehensive theoretical background is given. This is followed by a brief introduction of the most representative materials that exhibit polar nematic phases so far. Next, we expand on the experimental studies of polar phase, phase transitions, molecular dynamic simulations and anchoring and electrooptic studies of ferroelectric nematic materials. Finally, we highlight the most important challenges to be treated in future research.

## II. NEMATIC PHASES AND THEORETICAL BACKGROUND

### A. Nematic Phases

If we just focus on molecular liquid crystals, they can be classically divided into thermotropic and lyotropic LCs. While the latter refers to materials in which the liquid crystal phase transitions occur by variation of the concentration of the liquid crystal molecules in a solvent (most commonly water), for thermotropic liquid crystals, phases appear as a function of temperature. Although many different molecular structures form LC phases, they all share a common requirement which is an anisotropic molecular shape. Such an anisotropic shape will strongly determine the range of liquid crystalline phases exhibited by a material. There are multiple excellent texts covering the very various liquid crystalline phases and their properties [21–24], here we will exclusively focus on thermotropic nematic phases. For a long time, the range of known nematic phases was limited to a few, but the progressive investigation of new molecular geometries has led to the discovery of a new range of nematic mesophases. In this section, we shortly give a summary of them to establish the mesophase notation that will be followed throughout this contribution (Fig. 1). Non-chiral rod-like molecules can exhibit a uniaxial nematic phase, which is homogeneous (N). The N phase possesses inversion symmetry, and thus lacks polarity. The introduction of chirality in the rod-molecules, causes the director to twist in the chiral nematic LC (also known as cholesteric, N*). For odd liquid crystal dimers, the combination of molecular flexibility and bent-shape leads to the so-called twist-bend nematic phase ($N_{TB}$, Fig. 1), in which the director twists and bends in a heliconical structure. For bent molecules, also the splay counterpart, the so-called splay-bend nematic phase ($N_{SB}$, Fig. 1) has been predicted [1] involving a splay-bend modulation. These will be treated in detail in the next section. For biaxial constituents, the biaxial nematic phase $N_{bi}$ is predicted (Fig. 1) in which the directions perpendicular to the director are no longer degenerate. Despite the extended efforts, conclusive experimental discovery of the thermotropic biaxial nematic phase has been elusive so far [25–27].

The subject of this perspectives paper is the recently reported polar nematic phases, i.e., the ferroelectric nematic phase $N_F$, the ferroelectric chiral nematic phase $N_F$* and splay nematic $N_S$ phases. The $N_F$ phase, as in the case of the N phase, is uniaxial but inversion symmetry is broken, and the phase is polar, with the polarization along the director. The introduction of the chirality leads in this case to the $N_F$* in which both, the director and polarization twist along a helical axis (Fig. 1). On the other hand, the $N_S$ has been described as a modulated nematic structure, where splay deformation and polarization amplitude vary synchronously. Modulation can be 1D (Fig. 1) or 2D (see Fig. 1 in Ref. [28]).

Finally, it is worth noting here that thermotropic nematic materials can be combined with colloids to form interesting hybrid nematic systems with lower symmetry as for example biaxial [29], uniaxial or biaxial ferromagnetic nematic LC [30,31], or those exhibiting reconfigurable monoclinic colloidal nematic order [32].

### B. Macroscopic description

In this section, we will focus on several theoretical aspects required to build a macroscopic description of LC phases and relevant for the case of the ferroelectric nematic phases. Initially, a definition of the relevant order parameters is given, i.e. nematic order parameters, polarization and shape polarity. Then, Born´s approach for the description between a paraelectric isotropic and a ferroelectric nematic phase is presented. Finally, the fundamentals of the Landau-de Gennes approach for the description and analysis of phase transitions are given. A thorough introduction to order and phase transitions in soft materials, together with a comprehensive description of Landau and Landau-de Gennes approach to liquid crystals can be found in reference [33]. Here, the description of the Landau- de Gennes theory is accompanied by its application in several examples, starting from the simplest isotropic to nematic phase transition, and covering the different nematic-nematic transitions, from N-$N_{TB}$ to the present case of interest N-$N_S$ and N-$N_F$.

#### 1. *Order parameters*

An order parameter is defined as a physical quantity that is used to describe the order of a specific phase and is usually

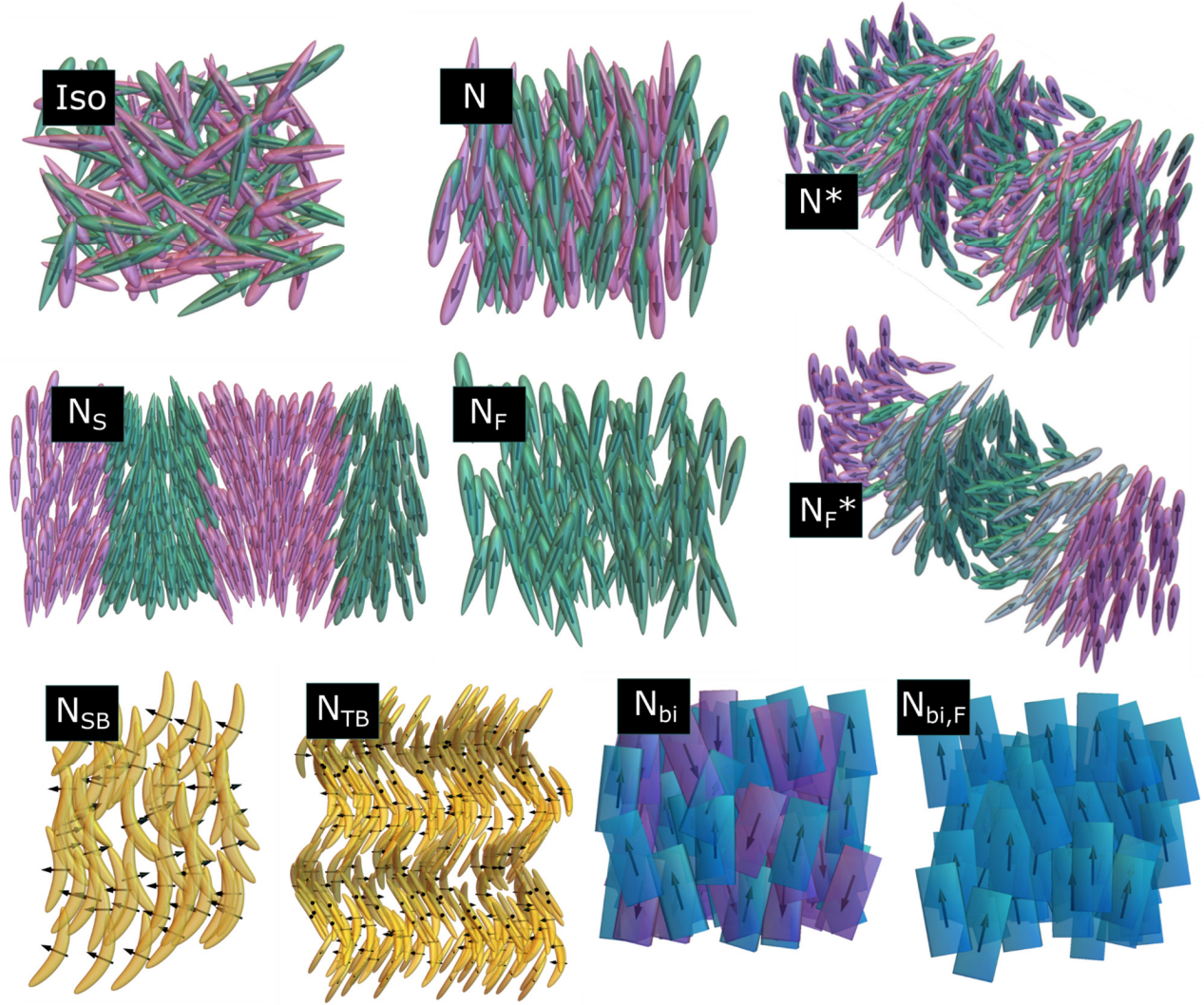

FIG.1. Schematic representation of isotropic phase (Iso) and different nematic structures, where arrow represent electric dipole of the constituent. For polar-shaped molecules, (N) nematic phase, (N*) chiral nematic or cholesteric phase, ($N_S$) splay nematic phase, ($N_F$) ferroelectric nematic pahse, ($N_F$*) chiral ferroelectric phase, ($N_{SB}$) splay-bend nematic phase and ($N_{TB}$) twist-bend nematic phase. For biaxial constituents, ($N_{bi}$) biaxial nematic phase and ($N_{bi,F}$) ferroelectric nematic phase. Here only cases with dipole moments along one symmetry axis are represented. In general, molecular dipole can be at an angle, which will contribute to the richness of possible phases, in particular in the case of biaxial nematics. Different colors are employed for the polar-shaped and biaxial constituents to indicate opposite dipole direction.

zero at the temperatures above the phase transition to that phase and has a finite value below it.

*(a) Order parameter of the nematic phase.* At the phase transition from isotropic to usual uniaxial nematic phase, the 2$^{nd}$ rank tensor quantities such as magnetic and electric susceptibilities become anisotropic. Thus, the order parameter of such uniaxial phase is a traceless 2$^{nd}$ rank tensor and can be written as $\mathbf{Q} = S\left(\mathbf{n} \otimes \mathbf{n} - \frac{1}{3}\underline{\mathbf{I}}\right)$. The unit vector **n**, called director, is oriented along the symmetry axis of the phase and has inversion symmetry, $\mathbf{n} \equiv -\mathbf{n}$. $S$ is the magnitude of the order parameter describing how well the constituents are on average oriented along **n**. The anisotropic parts of the magnetic and electric susceptibility tensors are proportional to $S$.

The anisotropic properties of the nematic phase are a consequence of the orientational order of constituents, e.g. molecules, which have anisotropic magnetic and electric polarizability tensors and/or are of anisotropic shape. In general, the orientation of a constituent is described by three Euler angles ($\alpha$, $\beta$, $\gamma$), and the probability for a given orientation is described by the orientational distribution function (ODF) $f(\alpha, \beta, \gamma)$.

In the uniaxial case (symmetry $C_\infty$), the ODF can be written as an even function of the cosine of the angle $\beta$

between the axis of a constituent and **n**. The orientation of a molecule can be defined in more ways. The obvious choices are the normalized eigenvectors of one of the polarizability tensors or the normalized eigenvectors of the inertia tensor. As the eigensystems of these tensors are in general not the same, the magnitude of $S$ calculated from the ODF as an average of the Legendre polynomial of the 2$^{nd}$ order $P_2$, $S = \langle P_2(\cos\beta)\rangle = \int f(\beta) P_2(\cos\beta) d(\cos\beta)$, will depend on this choice. This choice, however, will only determine the magnitude of the parameter $S$, while the terms and definitions of elastic constants remain the same.

In the case of a non-polar biaxial nematic, the set of possible choices of the order parameter is larger [34]. The 2$^{nd}$ rank tensor physical quantities also become biaxial, and in general, their eigensystems do not need to coincide. Unlike in the uniaxial case, there is no obvious best choice of the order parameter. If the eigensystems of the second rank tensors coincide, the biaxial symmetry is D$_{2h}$, else it is lower.

*(b) Polarization.* Polarization is defined as the volume density of molecular dipole moments and is a vector. Alternatively, the polar order can be described by a unit vector and an average of the Legendre polynomial of the 1$^{st}$ order $P_1$, $\langle P_1 \rangle = \langle \cos\beta \rangle = \int f(\beta)\cos\beta\, d(\cos\beta)$.

(c) *Shape polarity.* Materials made of constituents that lack head-tail symmetry can in principle also form a polar nematic phase even if they do not carry dipole moments. In this case, the physical quantity, which describes the orientation of the constituent is not a vector. The shape, described by the mass or volume distribution of a constituent can be expressed by a series of multipoles (or if the shape is not rotationally symmetric, in a series of Wigner rotation matrices *[35]*), similarly as is common for a charge distribution. The only difference regarding the charges is that there is no negative mass and, consequently, a mass dipole is zero. So the second non-zero multipole is the mass quadrupole (described by inertia tensor), which doesn't carry information about polarity. As already mentioned, it can be used for the description of the nonpolar nematic order. The lowest mass multipole that carries information about a head-tail non-symmetric shape is described by an octupole $T_{ijk}$, and this can be contracted to a vector $P_i = T_{ijj}$.

## 2. Born approach

As mentioned, Born [6] proposed a mean field model to explain the transition between paraelectric isotropic and ferroelectric nematic phases. In this model, the ferroelectric phase emerges as a consequence of dipolar interactions between the constituents. The material is assumed to be a dipolar liquid composed of molecules with dipole moments **p**. In an external electric field **E**$_0$, a given molecule is exposed to a local field, which is a sum of the external field and the field resulting from the polar order of the surrounding molecules. This polar order is described as a volume density of dipole moments, i.e. polarization **P**, and its contribution to the local field is proportional to it,

$$\mathbf{E}_{loc} = \mathbf{E}_0 + \alpha \mathbf{P} \qquad (1)$$

The resulting torque on a dipole causes that the molecules tend to orient along the local field, which can be described in terms of the potential energy,

$$-\mathbf{p}\cdot\mathbf{E}_{loc} = -pE_{loc}\cos\beta \qquad (2)$$

In a liquid, a molecule, and with it the dipole, can rotate. When the field is applied, it still rotates but on average it is oriented along the local field for longer time. The probability for a given orientation is given by the Boltzmann distribution $1/A \exp(-\mathbf{p}\cdot\mathbf{E}_{loc}/(k_B T))$, where $A$ is the normalization constant, $A = \int \exp(-\mathbf{p}\cdot\mathbf{E}_{loc}/(k_B T)) d\Omega$, and integration is performed over a solid angle $\Omega$. The magnitude of **P** can then be written as an average,

$$P = \rho_N p \langle \cos\beta \rangle \qquad (3)$$

Where $\rho_N$ is the number density, $\rho_N = \frac{N}{V}$. By defining $T_0 = \rho_N \alpha p^2/(3k_B)$ and $x = \alpha pP/(k_B T)$, Eq. (3) becomes

$$\frac{T}{3T_0} x = L(x) \qquad (4)$$

where $L(x)$ is the Langevin function, $L(x) = \coth(x) - 1/x$. This equation gives a real solution for $x$ when the temperature $T$ is lower than $T_0$. So this model predicts that at $T_0$, the system undergoes the transition to the polar phase and below this temperature, it exhibits spontaneous polarization $P = k_B T x_T/(\alpha p)$ where $x_T$ is the solution of Eq. (4) at temperature $T$.

In Eq. (1), the contribution $\alpha \mathbf{P}$ to the local field is a sum of the dipolar fields of the surrounding molecules. In the mean field approach, $\alpha$ is the depolarization factor, which for a spherical cavity is equal to $1/(3\varepsilon_0)$. This is the value that Born took. However, the magnitude of the dipolar field decreases as $r^{-3}$, so on a given molecule, the nearest neighbors are those that contribute the most to the field. Moreover, because of the anisotropy of the dipolar field, the local field will strongly depend on the relative positions and orientations of the neighbors. In magnetic dipolar liquids, the simple mean field approach has been shown to overestimate the local field and the question of whether the dipolar interaction alone can lead to polar order remained unanswered [36]. On the other hand, the shape of the constituents can strongly affect the positional and orientational correlations between neighbors, and thus, it can promote or prevent the formation of the polar phase. A dipolar liquid made of plate-like constituents is more likely to exhibit a polar nematic phase than the one made of rods [9,37].

Finally, it is important to remark here that, considering the current understanding, the polar nematic phases discussed in this paper are not the phases predicted by the Born model. This will be further elaborated in Section IV.C.

### 3. *Landau - de Gennes approach*

A general and very useful way to describe and analyze phase transitions (and general field dynamics) is via Landau-de Gennes (in general context Landau-Ginzburg) approach [38–40]. It consists of writing the free energy of a system in the high symmetry phase in terms of a characteristic quantity – order parameter. Such characteristic quantity is usually defined to be zero in the high symmetry phase and nonzero in the low symmetry phase. The free energy must be constructed in terms of the invariants (scalars) of the order parameter with respect to the high symmetry phase. The equilibrium state is then found by minimization of the free energy with respect to the order parameter.

Often there are other quantities that also appear in the low symmetry phase and are of interest. Such quantities are called secondary order parameters. The Landau - de Gennes (LdG) free energy then contains their invariants and the invariant coupling terms with the primary order parameter. In order to treat the possible deformations of the phase and to include the possibility of spatially modulated phases, gradient terms must be added to the free energy $F$.

Let us first briefly look at the basic LdG $F$ for the transition from the isotropic (I) to the nematic (N) phase as there are some issues that will be relevant for the properties of the polar N phases. The free energy density is [38]

$$f = \tfrac{1}{2} A \mathbf{Q}^2 + \tfrac{1}{3} B \mathbf{Q}^3 + \tfrac{1}{4} C \mathbf{Q}^4 + \tfrac{1}{2} L_1 Q_{ij,k} Q_{ij,k} + \tfrac{1}{2} L_2 Q_{ij,j} Q_{ik,k}$$
(5)

Here $Q_{ij,k} = \partial Q_{ij} / \partial x_k$. The quadratic coefficient $A$ is assumed to become negative at some temperature $T_c$ and so drives the transition. The cubic $B$ term causes the transition to be first order. B is also important in another respect. When $B = 0$, the low-temperature order is degenerate. The minimum of $f$ occurs for a manifold of states from oblate uniaxial with $S < 0$ to prolate uniaxial with $S > 0$ and biaxial states in between. If $B < 0$ the N phase is prolate uniaxial (usual calamitic N), and with $B > 0$ the low phase is oblate uniaxial (molecules are on average perpendicular to an axis but disordered in the plane). Therefore, the $B$ term forces the uniaxial state with an effective potential $|BS| p_b^2$, where $p_b$ is the degree of biaxial order. The empirical values for $B$ are quite large [41]. For example, a bend deformation by symmetry induces local biaxiality. By taking the value of $B$ from [41] and calculating $p_b$ in a hybrid cell (with planar boundary conditions on one surface and homeotropic on the other) we get that biaxiality becomes appreciable only when the bend radius is of the order of 1 nm.

The second point concerns the elastic terms $L_1$ and $L_2$. The elastic energy of a nematic is usually written in terms of the director. It has the form, first proposed by Frank and Oseen,

$$f_N = \tfrac{1}{2} \left[ K_1 (\nabla \cdot \mathbf{n})^2 + K_2 (\mathbf{n} \cdot (\nabla \times \mathbf{n}))^2 + K_3 (\mathbf{n} \times (\nabla \times \mathbf{n}))^2 \right]$$
(6)

The first term is the splay deformation, in which the director diverges or converges, the second is the twist (perpendicular to the director) and the third term corresponds to the bend deformation. The Frank-Osen elastic energy contains three elastic constants, splay ($K_1$), twist ($K_2$), and bend ($K_3$). In terms of $L_1$ and $L_2$, $K_1 = K_3$. This degeneracy is broken by higher-order terms of the form $\mathbf{Q}(\nabla \mathbf{Q})^2$ [42,43].

We also need to consider the polarization vector $\mathbf{P}$. This does not need to be electric polarization but can also be steric, describing for example pear-shaped molecules. Polarization brings new terms to the free energy including the couplings to the nematic order. The lowest term $\mathbf{P} \cdot \mathbf{Q} \cdot \mathbf{P}$ forces $\mathbf{P}$ to be parallel to the largest or smallest eigenvector of $\mathbf{Q}$. In the usual uniaxial case, this means that $\mathbf{P}$ is either parallel or perpendicular to $\mathbf{n}$, depending on the sign of the term.

The second very important terms are the so called flexoelectric couplings of $\mathbf{P}$ to $\nabla \mathbf{Q}$. Their importance was first recognized by Frank [7] and later by R. Meyer [44] who also recognized that flexoelectricity can lead to instabilities. In terms of the director, there are two such terms: splay $e_1 \mathbf{n} (\nabla \cdot \mathbf{n}) \cdot \mathbf{P}$ and bend $e_3 (\mathbf{n} \times \nabla \times \mathbf{n}) \cdot \mathbf{P}$. To the lowest order in $\mathbf{Q}$, there is just one flexoelectric term $(\nabla \cdot \mathbf{Q}) \cdot \mathbf{P}$. So, as in the case of elastic constants, the flexoelectric coefficients are also degenerate to this order. Such degeneracy is then broken by terms of the form $P_i Q_{ij} Q_{jk,k}$ ($\mathbf{P} \cdot \mathbf{Q} \cdot \nabla \cdot \mathbf{Q}$). We note here that flexoelectric coupling, together with the term $\mathbf{P} \cdot \mathbf{Q} \cdot \mathbf{P}$, also breaks the degeneracy of $K_1$ and $K_3$.

Considering a polar nematic phase, one can take just $\mathbf{P}$ as the primary order parameter. This was done by Pleiner and Brand in 1989 [8]. They wrote the free energy density in the form

$$f = -\tfrac{1}{2} a \mathbf{P}^2 + \tfrac{1}{4} b \mathbf{P}^4 - c_0 \nabla \cdot \mathbf{P} + \tfrac{1}{2} c_1 (\nabla \cdot \mathbf{P})^2 + \tfrac{1}{2} c_2 (\nabla \times \mathbf{P})^2$$
(7)

As $\mathbf{P}$ is a polar vector, the term $c_0$ div $\mathbf{P}$, which is linear in splay deformation, is allowed and leads to a structure with spontaneous splay. Since it is impossible to fill the space with homogeneous splay, the authors discussed possible structures that included defects.

*(a) $N_{TB}$ and $N_{SB}$ phases*

With a free energy in terms of $\mathbf{n}$, including second-order gradient terms, I. Dozov considered the case where the bend elastic constant $K_3$ would become negative so that the homogeneous N phase becomes unstable [1]. As homogenous bend also cannot fill the space, the new phase can exhibit either twist-bend ($N_{TB}$) or splay-bend ($N_{SB}$) modulation (Fig. 1), depending on the ratio $K_1/K_3$. The $N_{TB}$ phase was indeed found by M. Cestari et al [3], V. Borshch [5] and D. Chen [4]. In Ref [45], it was also shown by light scattering that indeed $K_3$ becomes very small close to the transition.

Following the observation of Meyer [44] that a bend must be accompanied by flexoelectric polarization, Shamid et al. [46] proposed a model for the transition from N to the $N_{TB}$ phase. They started with a free energy density

$$f = \tfrac{1}{2}K_1\mathbf{S}^2 + \tfrac{1}{2}K_2 T_{tw}^2 + \tfrac{1}{2}K_3\mathbf{B}^2 + \tfrac{1}{2}\mu\mathbf{P}^2 - \lambda\mathbf{B}\cdot\mathbf{P} + \\ + \tfrac{1}{4}\nu\mathbf{P}^4 + \tfrac{1}{2}\kappa(\nabla\mathbf{P})^2 \qquad (8)$$

where $\mathbf{S}$, $T_{tw}$, and $\mathbf{B}$ denote the splay, twist, and bend deformations. The flexoelectric coupling renormalizes the bend elastic constant to $K_{3eff} = K_3 - \lambda^2/\mu$. This shows that the measured $K_{3eff}$ can become 0 due to either decreasing $K_3$ (bare elastic constant) or $\mu$. In Ref [46], $\mu$ is taken to be temperature dependent, so $\mathbf{P}$ is the primary order parameter and drives the transition. In the $N_{TB}$ phase, $\mathbf{P}$ forms a helical structure and thus, the $N_{TB}$ can also be considered as a ferrielectrical phase. The paper also finds the same conditions as Dozov for either $N_{TB}$ or $N_{SB}$ to be stable. Additionally, it presents numerical simulation results of a lattice model that also shows a periodic TB or SB structure.

Shamid et al. [47] looked at the possibility that polarity of the N phase would result in hexagonal or cubic periodic structures with twist and bend, in an analogous way to the chirality induced blue phases. In addition to a numerical lattice calculation, they present an LdG analysis based on the free energy

$$f = \tfrac{1}{2}A\mathbf{Q}^2 + \tfrac{1}{3}B\mathbf{Q}^3 + \tfrac{1}{4}C\mathbf{Q}^4 + \tfrac{1}{2}L_1 Q_{ij,k}Q_{ij,k} + \tfrac{1}{2}L_2 Q_{ij,j}Q_{ik,k} + \\ + \tfrac{1}{2}\mu\mathbf{P}^2 + \tfrac{1}{4}\nu\mathbf{P}^4 + \tfrac{1}{2}\kappa(\nabla\mathbf{P})^2 + \lambda\mathbf{P}\cdot(\nabla\cdot\mathbf{Q}) + \eta\mathbf{P}\cdot\mathbf{Q}\cdot\mathbf{P} \qquad (9)$$

where $\mathbf{Q}$ and $\mathbf{P}$ are expanded in the lowest order Fourier modes. The model predicts, besides I, N, and $N_{TB}$ phases, also a stable cubic BCC and a 2D hexagonal phase consisting of a lattice of defects. The latter phases have not been observed yet.

Longa and Pajak [48] analyzed the possible 1D periodic structures by including both $\mathbf{P}$ and intrinsic chirality. To Eq. (9) they added a chiral term $\mathbf{Q}\cdot(\nabla\times\mathbf{Q})$, which generates the $N^*$ phase, and a chiral coupling $(\nabla\cdot\mathbf{Q})\cdot(\nabla\times\mathbf{P})$. This last term is not allowed for electrical polarization but is possible for steric dipoles. Two fourth order terms were also added. With all, six possible structures were found including the usual $N^*$ phase. All phases except $N^*$ have also periodically modulated $\mathbf{P}$ and biaxiality. The $N_S$ phase is not among those structures. In both papers [47,48], $K_1$ and $K_3$ are degenerate and the $N_{TB}$ phase is obtained by the positive $\mathbf{P}\cdot\mathbf{Q}\cdot\mathbf{P}$ term favoring the bend flexopolarization ($\mathbf{n}$ perpendicular to $\mathbf{P}$).

*(b) $N_S$ and $N_F$ phases*

We now move to the main theme of this paper, which is the splay $N_S$ (Fig. 1) and the homogenous ferroelectric $N_F$ phase (Fig. 1). In Ref [18], it was shown by dynamic light scattering that in the upper N phase of RM734 the splay constant $K_1$ becomes very small before the transition to the $N_S$ phase. The transition is modeled by a free energy

$$f = \tfrac{1}{2}L_1 Q_{ij,k}Q_{ij,k} - L_5 Q_{ij}Q_{ik,k}Q_{jl,l} - \gamma P_i Q_{ij,j} + \tfrac{1}{2}t\mathbf{P}^2 + \tfrac{1}{2}b(\nabla\mathbf{P})^2 \qquad (10)$$

This is the simplest elastic energy where $K_1 = L_1 S^2 - 2/3 L_5 S^3$ and $K_3 = L_1 S^2 + 4/3 L_5 S^3$ are not degenerate. The term $(\nabla\mathbf{P})^2$ is necessary to stabilize the splayed state. Assuming uniaxial symmetry, the results are analogous to that in Ref [46]. The transition to $N_S$ occurs when $K_{1eff} = L_1 S^2 - 2/3 L_5 S^3 - \gamma^2/t = 0$. The bend constant is also renormalized: $K_{3eff} = L_1 S^2 + 4/3 L_5 S^3 - \gamma^2/t$, but if $L_5 > 0$, splay becomes unstable first. The instability can be driven either by decreasing bare $K_1$ due to the dependence on $S$ or by decreasing $t$, that is by the ordering of $\mathbf{P}$. The form of the $T$ dependence of $K_{1eff}$, shown in Fig. 6(a), where $K_{1eff}$ is only slowly decreasing to about 5 K above the transition and then rapidly goes to 0, is consistent with $t$ driving the transition. The model (5) also predicts that close below the transition temperature $t_c = \gamma^2/K_1$ the amplitude of the splay angle and the modulation wavevector are proportional to $\sqrt{t_c - t}$ and $\mathbf{P}$ to $(t_c - t)$.

In Ref [18], it is also pointed out that if $L_5 < 0$, free energy (5) predicts the $N_{TB}$ phase. This is further elaborated in Ref. [49]. The unified treatment of $N_s$ and $N_{TB}$ phases is one of the advantages of using the tensor order parameter $\mathbf{Q}$ to construct the free energy. The dependence of the $L_5$ term on $S^3$ could also be important in driving the $N_{TB}$ transition, as suggested in Fig. 2 of [49]. By parametrizing $\mathbf{Q}$ as a biaxial tensor, Eq. (10) would also give the amount of local biaxiality due to splay or bend. In that case, it is necessary to include the $B\mathbf{Q}^3$ term, as discussed above and done in [48].

Chaturvedi and Kamien [50] analyzed the mechanisms for the transition to the modulated N phases in terms of $\mathbf{n}$. They tried bond orientational ordering and found that in the lowest order it does not lead to the modulated phases. They conclude that the flexoelectric coupling is the important mechanism and consider the symmetry of the $N_{SB}$ and $N_S$ phases on exchanging bend and splay flexoelectric coupling.

The periodicity of the splay in $N_S$ can also be 2D. This was pointed out by Rosseto and Selinger [28]. They considered the free energy in Eq. (9) and expanded it to fourth order in all variables. They found that doubly periodic splay has slightly lower energy except very close to the $N_S$ transition and $K_1/K_3 > 3$. They also found that, at a scaled temperature sufficiently below the transition, the homogeneous ferroelectric phase $N_F$ is the stable one.

The stability of the $N_F$ phase was also studied by E. Katz [51]. He takes free energy with $\mathbf{P}$ as the primary parameter:

$$f = \tfrac{1}{2}t\mathbf{P}^2 + \tfrac{1}{4}\lambda\mathbf{P}^4 + \tfrac{1}{2}b(\nabla\mathbf{P})^2 + \beta_1\mathbf{P}^2(\nabla\cdot\mathbf{P}) + \beta_2(\mathbf{P}\cdot\nabla)\mathbf{P}^2 +$$
$$+ \tfrac{1}{2}K(\nabla\mathbf{n})^2 + \gamma(\mathbf{P}\cdot\mathbf{n})^2 + e_1\mathbf{P}\cdot\mathbf{n}(\nabla\cdot\mathbf{n}) + e_2\mathbf{P}\cdot(\mathbf{n}\cdot\nabla)\mathbf{n}$$

(11)

where one nematic elastic constant approximation is used. "Flexodipolar" terms of **P** coupled to $\nabla\mathbf{P}$ are also added. Depending on the sign of γ, **P** is either perpendicular (γ > 0) or parallel γ < 0) to **n**. In both cases, the condition of stability is derived by considering small perturbations of the ground state. In the perpendicular case, it is found that $N_F$ is stable if the flexodipolar interaction is neglected (β = 0). In the parallel case (the experimentally observed $N_F$) the stability criterion is somewhat complicated and depends on all the parameters. The instability occurs for the 1D and 2D periodic perturbation at the same point in parameter space.

In discussing the stability of the homogeneous $N_F$ phase it is necessary to note a paper by A. G. Khachaturyan [52]. In this paper, the author considers the electrostatic self energy of **P**, which is not included in the models above. The result points out that a ferroelectric fluid is unstable against a twist with an axis perpendicular to **P**, analogous to the N* phase.

*4. Electrostatics*

In the description of ferroelectric material, there is an additional contribution to the energy of the system due to the so called depolarization field, which is the field created by the spontaneous electric polarization. In the absence of free charges and external field, it can be calculated using Gauss's law $\nabla\cdot\mathbf{D} = \nabla\cdot(\varepsilon_0\boldsymbol{\varepsilon}\mathbf{E}_{dep} + \mathbf{P}) = 0$, where **ε** is a dielectric tensor. Effectively this means that the depolarization field $\mathbf{E}_{dep}$ is generated by "bound charges" generated by $-\nabla\cdot\mathbf{P}$. For a finite piece of material, they can be divided into surface charge $\sigma_b = \mathbf{v}\cdot\mathbf{P}$, where **v** is normal to the surface, and volume charge $\rho_b = -\nabla\cdot\mathbf{P}$. These charges are sources of electrostatic potential from which $\mathbf{E}_{dep}$ is calculated. The contribution of the depolarization field to free energy is

$$F_{dep} = -\tfrac{1}{2}\int_V \mathbf{E}_{dep}\cdot\mathbf{P}\,dV \quad (12)$$

where integration is performed over the ferroelectric body. As the ferroelectric NLCs exhibit large polarization values, this contribution to the total free energy is important. In the general case, the calculation of the structure affected by the depolarization field is challenging. However, if **P** is parallel to **n**, i.e., $\mathbf{P} = P\mathbf{n}$, and it depends only on one Cartesian coordinate (let's say *z*), for example, when in a thin layer oriented perpendicularly to *z* and confined by non-conducting boundaries, then the depolarization field energy becomes

$$F_{dep} = \tfrac{1}{2}A\int_0^d \frac{n_z^2(z)P^2(z)}{\varepsilon_0(\Delta\varepsilon n_z^2(z) + \varepsilon_\perp)}dz \quad (13)$$

where $n_z$ is the *z*-component of **n**, $\Delta\varepsilon = \varepsilon_\parallel - \varepsilon_\perp$ with $\varepsilon_\parallel$ and $\varepsilon_\perp$ being the components of **ε** parallel and perpendicular to **n**, and *A* the surface of the layer. From Eq. (13), it follows that the depolarization energy favours that **P** lays in the plane of the layer and that the field of the order of $E_{ext} \approx P/(\varepsilon\varepsilon_0)$ is needed to orient **P** perpendicularly to the layer. Similarly, when the system exhibit 1D modulated splay structure, the integrand in Eq. (13) can be added as a new term to free energy density in Eq.(10). For small splay modulations, for example at the phase transition from N to $N_s$ phase, this term only slightly rescales the $\mathbf{P}^2$ term and causes a small reduction of $t_c$.

In a general case, when **P** (and **n**) depends on more coordinates, the material is surrounded by conducting homogenous or patterned surface, e.g. electrodes, or/and the ions are present, the combined LdG and Poisson-Boltzmann theoretical framework (e.g. as in Refs. [53,54]) should be used for the studies of possible structures.

### C. Difference with other polar liquid crystalline mesophases

It is important to recall here, the differences between the ferroelectric nematic phases subject of this manuscript and classical ferroelectric liquid crystals. Here we are dealing with LC phases without any degree of positional ordering. In the LC community, the nomenclature ferroelectric liquid crystals was historically employed to address tilted layered mesophases shown by chiral mesogens (SmC*), in which molecules are tilted with respect to the layer normal and due to the molecular chirality such tilt turns from layer to layer forming an helix. As pointed out by R. B. Meyer [55] if there is a component of the molecular dipole perpendicular to the molecular long axis, symmetry allows the existence of spontaneous polarization in the direction perpendicular to the tilt plane. As the tilt plane turns around the layer normal from layer to layer, the direction of the layer polarization also makes the same turn. Although the SmC* is truly defined as helielectric, it is usually labeled as ferroelectric because of the ferroelectric phase generated by unwinding the helix. Opposite tilts in adjacent layers give rise to the antiferroelectric counterpart $SmC_A$*. Nice reviews about classical ferroelectric mesophases can be found in references [56–58].

In 1996 Niori et al. [59] described a new system of polar mesophases formed by non-chiral rigid bent-core molecules, where the layer normal, the tilt direction and the layer polarity define a system which is either left-handed or right-handed. Then, the bent-core molecules can develop chiral structures and with them polar phases, despite being achiral. In contrast to polar phases exhibited by calamitics, in bent-core molecules polarization and tilt are no longer correlated. A change of the tilt or polarization direction implies the inversion of the layer chirality. Numerous layered polar structures have been described for bent-core molecules for which nice reviews can be found in references [25,60,61].

Also worth mentioning are efforts devoted to promote polar columnar phases, consisting of disk-shaped molecules in which molecules stack to form columns [62–64], showing polar order with polarization along the columns. For a detailed review of such phases, readers can address to reference [65].

This section does not aim to be an in-depth review of ferroelectricity in liquid crystalline phases but to highlight the differences with respect to the newly described polar nematic phases. The different systems described in this section, possess some degree of long-range positional order as a requisite for the appearance of polar ordering. However, as pointed out before, polar nematic phases are unique, in that only long-range orientational order is present and 3D fluidity is retained.

## III. MATERIALS

In this chapter, the materials which have been shown to exhibit ferroelectric nematic phases so far will be briefly presented.

In Fig. 2, three representative materials exhibiting the ferroelectric nematic phase are shown. All three have a large longitudinal dipole moment component. DIO and the material **1** from reference [66] belong to fluorinated nematogen materials, which are widely used in liquid crystal displays [67]. Although this kind of materials has been extensively studied in the past, so far, materials similar to DIO that show the ferroelectric nematic phase are scarce [68]. On contrary, Mandle *et al* already showed that several materials similar to RM734 exhibit two nematic phases [16,17]. Besides large longitudinal dipole moment, these materials have a terminal nitro group and lateral alkyloxy group. The appearance of lower temperature, ferroelectric nematic phase is strongly sensitive to small structural changes, e.g. replacing the terminal nitro group with another dipole group, removal of a single ester unit or the lateral group from RM734 prevents its formation [16,69]. On the other hand, moving the lateral group to the middle ring or fluorination of any of the rings affects the phase transition temperatures and, in some cases, increases the ferroelectric polarization [16,68,70]. The obvious question is then what molecular features are favorable for the appearance and stability of the polar nematic phases. J. Li et al. reported a series of molecules starting from close derivatives of DIO and RM734 and subsequently shortening or lengthening the molecules, introducing ester, fluoride, fluorocarbon ether, nitro, nitrile or trifluorocarbon groups as electron-withdrawing groups to control the molecular dipole moment [68]. Materials are then classified according to the exhibit of stable, metastable or absent polar nematic phase. The authors conclude that, statistically, the most critical parameter is the magnitude of the molecular dipole moment ($\mu > 9D$).

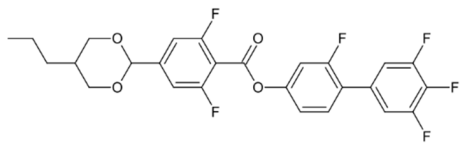

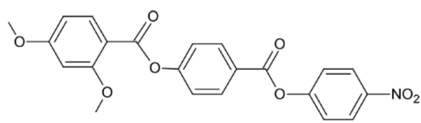

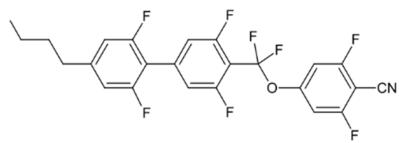

FIG.2. Chemical structure of three representative materials exhibiting ferroelectric nematic phase and their phase sequence.

The third molecule represented in Fig. 2, material **1** from A. Manabe et al. in reference [66], contains fluorinated biphenyl- and phenyl units bridged by a difluoromethoxy group and is terminated by a nitrile group [66]. Note, that contrary to RM734 and homologues, nitrile group does not hinder in this case the appearance of the polar phase. The calculated dipole moment adds up to approximately 11.3 D. Remarkably, material **1** from reference [66] shows direct Iso-$N_F$ transition, being the ferroelectric phase accessible at temperatures slightly below room temperature. It is speculated that the relatively long butyl chain is responsible in this case for the low temperatures.

The polar phase in low molecular weight RM734-like materials has been shown to extend also to rigid oligomers [71] by J. Li et al. and polymers [68,72] by S. Dai et al.. Oligomers made of repeating p-oxybenzoate units with the propoxy side group (Fig. 3) have dipole moments, which increase with the number of repeated units and can reach values over 30 D. Most of them exhibit a direct transition from isotropic to ferroelectric nematic phase [71], with transitional temperatures shifting up with the increase of the oligomer length.

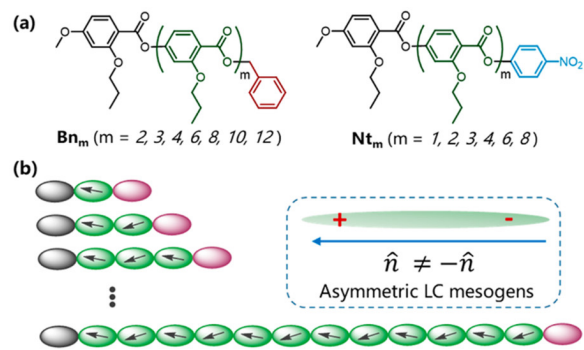

FIG.3. Oligomers. (a) Chemical structure of $Bn_m$ and $Nt_m$ main-chain LC oligomers. (b) Schematic representation of the mesogens. © 2021, American Chemical Society. Reprinted with permission, Ref. [71].

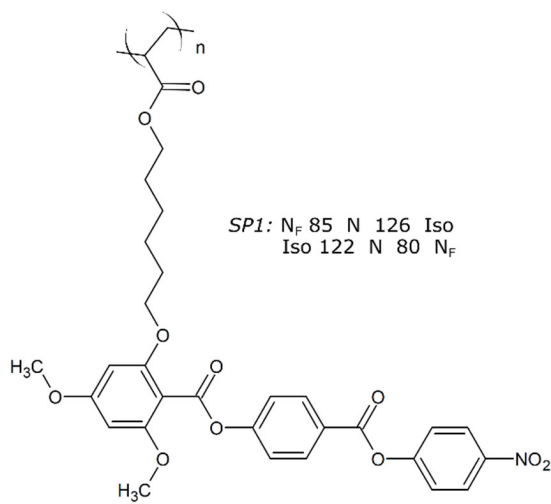

FIG.4. Chemical structure of the SP1 side-chain liquid crystalline polymer and its phase sequence on heating and cooling. Ref. [72].

On the other hand, Dai e*t al* [68,72] studied three side-chain LC polymers based on RM734 and showed that liquid crystal polymers with RM734-like mesogen side-joined (Fig. 4) form the polar phase, while the polymers with end-jointed mesogens self-assemble into a smectic A phase with no polarity. They also showed that the magnitude of the dipole moment of the side mesogen is critical for the appearance of macroscopic polarity in these liquid crystalline polymers. In this case, transition temperatures are also quite above room temperature, with melting temperatures above 100 °C.

As shown in this chapter, despite the early stages of the research, in the past few years a number of materials showing the polar nematic phases have slowly accumulated.

## IV. POLAR PHASES AND PHASE TRANSITIONS

In this Chapter, the most significant current experimental findings will be covered. An introduction of methods and their challenges in the frame of research on ferroelectric nematic materials will be presented in Chapter VI.

### A. Non-chiral phases and phase transitions

As mentioned above, three distinct nematic phases were originally reported for DIO by Nishikawa et al. [15]. On cooling from the higher temperature mesophases DSC curve reveals two exothermic peaks at 84.5 °C ($\Delta H$ =0.003 kJ mol$^{-1}$) and 68.8 °C ($\Delta H$ =0.2 kJ mol$^{-1}$), evidencing two weak first order transitions between three mesophases (see Fig. 5). As will be discussed below, those phases can be identified, from high to low temperature as N (M1), $N_S$ (M2), and $N_F$ (MP). The nomenclature N, $N_S$, $N_F$ will be used from now on for clarity. The identification of M2 as $N_S$ to designate the intermediate phase found in DIO will be discussed in detail in section VI. It is worth recalling here the much weaker character of the N- $N_S$ transition when compared to $N_S$-$N_F$ one. Further cooling results in the crystallization of the sample at 34 °C.

The same year, Mandle et al. reported for RM734 (Compound 2 in reference [16]) the phase sequence Cr 139.8 °C ($N_x$ 132.7 °C) N 187.9 Iso, where $N_x$ is now known to be $N_F$. Initially labelled as $N_S$, the discussion about the nomenclature is left for Section VI.B. The associated enthalpy of the N-$N_F$ transition is in this case also very small 0.2 kJ mol$^{-1}$, comparable to that of $N_S$ -$N_F$ measured for DIO. Similar values are reported for RM734 homologues [16,73].

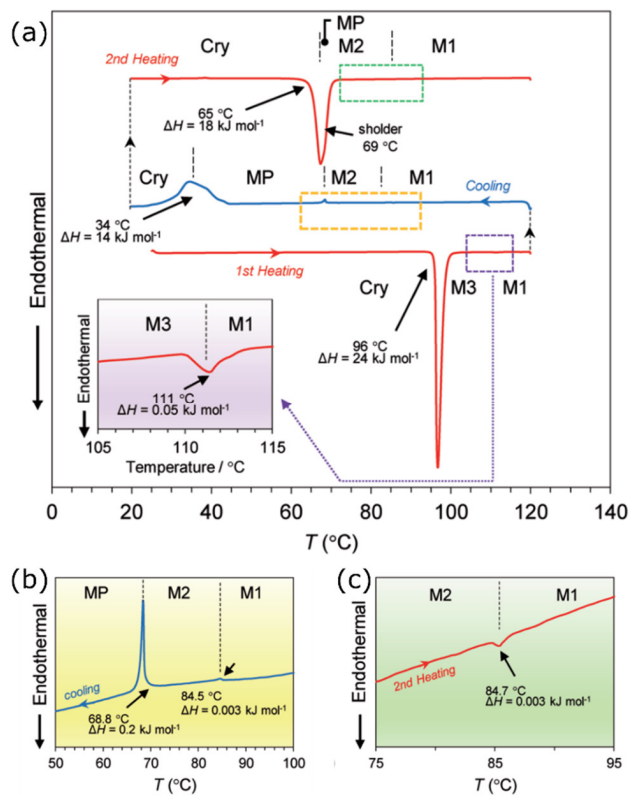

FIG.5. DSC curves of DIO at 5.0 K/min for heating and cooling scans (a) First heating scan (bottom) and subsequent cooling and heating runs. (b) Zoom in detail of M1-M2 (N-$N_S$) and M2-MP ($N_S$-$N_F$) transitions on cooling. (c) A close look at the M2-M1 ($N_S$-N) transition during second heating. © 2017 WILEY-VCH Verlag GmbH & Co. KGaA, Weinheim. Reprinted with permission, Ref. [15].

Recently, by means of precision adiabatic scanning calorimetry, J. Thoen et al. have shown that the N-$N_F$ transition in RM734 is close to a tricritical point [74].

Despite coming from different molecular families, X. Chen et al. have recently shown that DIO and RM734 show complete miscibility [75]. All the mixtures show Iso, N, and $N_F$ phases. Such results importantly demonstrate that the low temperature phases of both materials are the same phase. The reported mixtures also show the appearance of the $N_S$ phase for DIO concentrations above 50%. For intermediate concentrations, the poor compatibility of the crystal phases of both materials results in the suppression of the crystallization and the stabilization of the ferroelectric nematic phase at lower temperatures.

*1. X-ray scattering and birefringence*

In either case, DIO or RM734, X-ray experiments [15,16,76] show only diffuse scattering at both small and wide angles, and thus the presence of long-range translational order was ruled out and the mesophases were assigned an N-like character. Additionally, a couple of distinctive characteristics of the X-ray behavior have been highlighted in the case of RM734, with respect to classical nematic liquid crystals [16]. In either of the fluid phases, the scattering intensity is extremely weak, requiring long exposure times to obtain good signal-to-noise ratios. Besides, in both nematic phases (N and $N_F$) additional diffuse peaks (002, 003) are seen at small angles parallel to the external aligning field, i.e. along the director, Fig. 6(c). Such observations have also been reported for different members of the RM734 family, all of them exhibiting the $N_F$ phase [17,70]. Similar features have also been described for DIO [75]. Interestingly, for an RM734 homolog material in which the nitro group is replaced by a nitrile group (RM734-CN, which uniquely exhibits N phase) none of these behaviors is observed [70,76]. As will be discussed below in the MD simulation section, such additional small-angle scattering peaks seem to be the result of polar ordering in the polar state [70].

Orientational order parameters $\langle P_2 \rangle$, $\langle P_4 \rangle$, $\langle P_6 \rangle$ and $\langle P_8 \rangle$ have been extracted from SAXS/WAXS measurements for RM734 and show a discontinuous increase at the N-$N_F$ transition, while $\langle P_{10} \rangle$ is effectively zero at any temperature (see Fig. 6(e)) [76]. The increase of orientational order at the transition is also evidenced in the birefringence. Measurements of $\Delta n$ for both materials show similar temperature behaviour, with an increasing value of $\Delta n$ while decreasing temperature in the N phase. The N-$N_F$ transition is then characterized by the jump of $\Delta n$ (see Fig. 6(g), Fig. S17 in Ref. [20] or Fig. (3)a for an RM734 fluorinated homolog in Ref. [73]). Such a sudden increase is not observed for DIO at the N-$N_S$ transition, where only a change in the slope is observed for $\Delta n$ [73] as shown by S. Brown

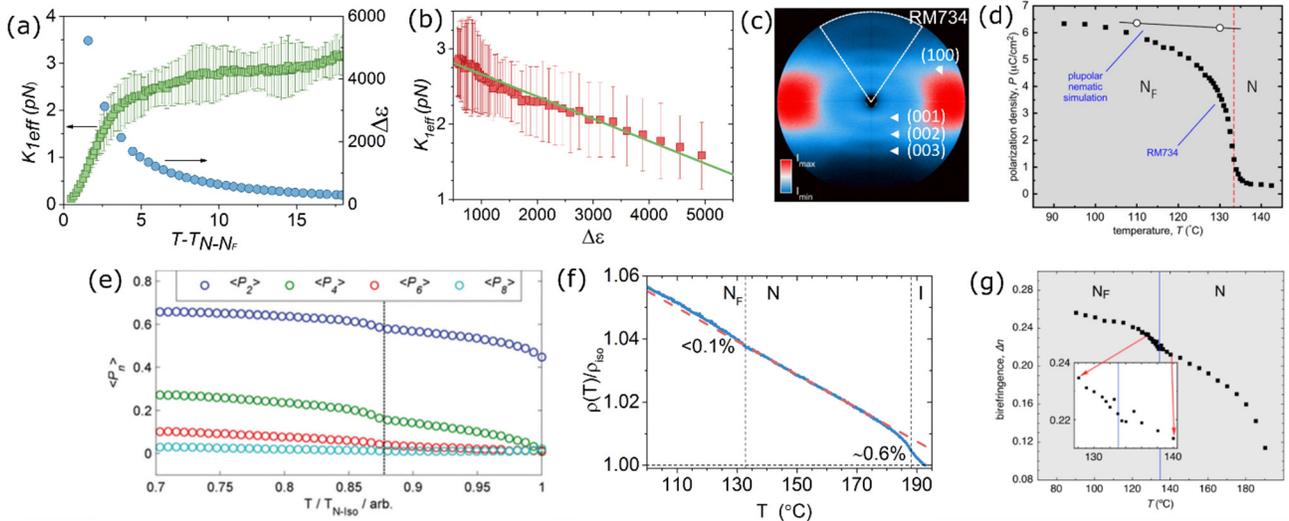

FIG.6. a) Temperature dependence of the effective splay elastic constant ($K_{1,eff}$) and amplitude of the main dielectric relaxation mode in the N phase of RM734 ($\Delta\varepsilon_{low}$). b) $K_{1,eff}$ vs interpolated values of $\Delta\varepsilon_{low}$ showing linear dependence. (a-b) © 2020 American Physical Society. Reproduced with permission, Ref. [19] c) Magnetically aligned two-dimensional X-ray scattering pattern in the N phase of RM734. Reproduced from Ref [70] licensed under CC BY 4.0 d) Temperature dependence of the electric polarization, P. Reproduced from [20], licensed under CC BY 4.0. e) Temperature dependence of the first four even order parameters Reproduced from Ref. [76], licensed under CC BY 3.0. f) Measured temperature dependence of the normalized density, where dashed lines is the extrapolated value form the N phase. Reproduced from Ref [70] licensed under CC BY 4.0. g) Temperature dependence across the N and $N_F$ phase of the birefringence. Reproduced from [20], licensed under CC BY 4.0

et al.. A jump in the birefringence is then observed at the $N_S$-$N_F$ transition.

*2. Transitions by polarizing optical microscopy*

Optical texture changes over the transitions have been extensively described for RM734 [18–20,77,78] and DIO [15,68,73,75]. For the latter, the N- $N_S$ transition is characterized by increased pretransitional nematic flickering, directly related to the fluctuations of the nematic director. At the transition, freezing of the strong fluctuating mode takes place which is observed as a wavefront propagating throughout the cell [73] as also pointed out by S. Brown et al., see Supplementary Movie 1 [79]. Textures of N and $N_S$ are similar, with that of the latter characterized by strongly reduced flickering and by the appearance of zig-zag defects (in cells with planar alignment, see Fig. 7) which slowly disappear on further cooling. Transition to the $N_F$ phase is then marked by the strong destabilization of the uniform texture in the form of stripes, which at a slightly lower temperature disappear and lead to a homogeneous texture equivalent to that in the N phase. This transition is also accompanied by material flow. It should be noted here that, as will be covered below in section V.A, the nature of the confining surfaces (alignment layer, rubbing, pretilt) exerts a strong influence on the final observed structure [77,78,80]. In very similar terms, in the case of RM734 (Fig. 7), the transition has been described to show strong pretransitional behavior in the N phase characterized by strong fluctuations followed by the freezing of the nematic flickering and immediately followed by destabilization of the homogenous orientation of the director, characterized by a striped texture [18,19,77]. This destabilization leads to the recovery of a homogeneous texture and with it the flickering. Transition on slow cooling can be found in Supplementary Movie 2 [81]. Comparing both materials and the description of the transitions, it can be conjectured that the narrow temperature range in which RM734 shows freezing of the nematic fluctuations indeed resembles a very narrow $N_S$ phase followed by a transition to the $N_F$ phase. We will return to this discussion in Section VI.B.

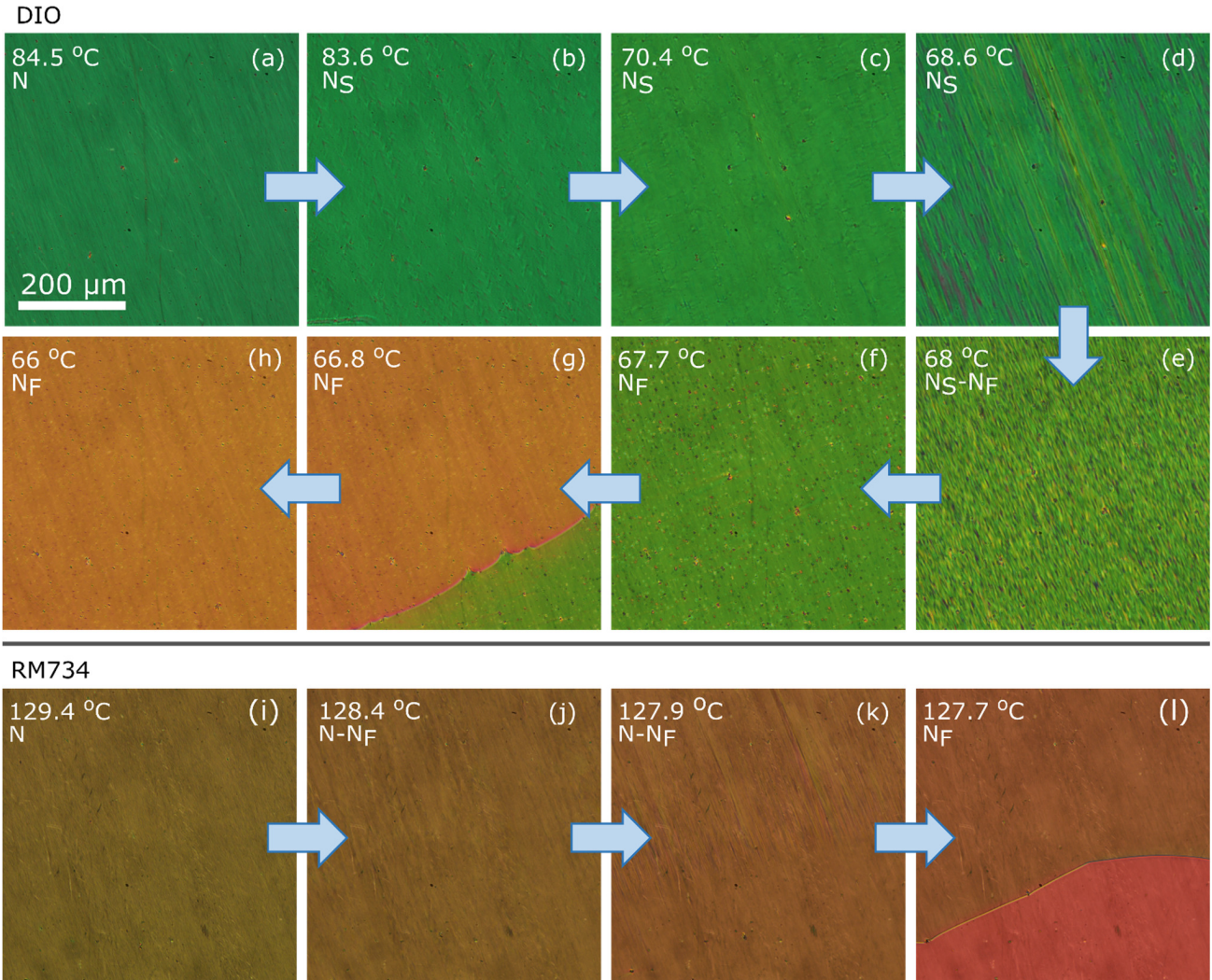

FIG.7. Polarizing optical microscopy images of the textures of DIO (a-h) and RM734 (i-l) on cooling through the N- $N_S$ -$N_F$ and N-$N_F$ phase sequences respectively. Images taken in EHC 8 μm planar cells with antiparallel rubbing. Image width 540 μm.

### 3. Pretransitional behaviour: $K_1$ and $\Delta\epsilon$

Studies of the director orientational fluctuations close to the phase transition in RM734 by A. Mertelj et al. show that the observed strong pretransitional fluctuations in the N phase are a consequence of the strong softening of the splay elastic constant $K_1$ [18,70]. $K_1$ is unusually low in the studied temperature range, and decreases towards zero when approaching the transition (see Fig. 6(a)). Decreasing $K_1$ has been also observed for DIO and several mixtures in the proximity of N-$N_S$ transition [75,82]. Interestingly, a comparison with RM734-CN shows that in the absence of the polar nematic phases, the $K_1$ value remains small, but the softening of the pretransition behaviour is absent [70].

Such pretransitional behavior is also reflected in the dielectric properties of RM734 in the preceding N phase. Measurements of the parallel component of the dielectric spectra for RM734 have been reported for a broad frequency range by N. Sebastián et al. [19]. The spectrum in the N phase is characterized by the presence of two low frequency relaxation modes (low frequency as lower frequency than in the isotropic phase), instead of the single relaxation found in classical nematic phases corresponding to the molecular reorientation around the long molecular axis (see Fig. 3 in reference [19]). Of both, that of lower frequency shows strong softening on approaching the N-$N_F$ transition, characterized by the fast decrease of its frequency, deviating from Arrhenius behavior, and the divergence of its amplitude, reaching values up to $\Delta\varepsilon_{low} \sim 6000$ (Fig. 6(a)). Together with the behavior of the splay elastic constant, it shows that the transition between nematic phases in RM734 is a ferroelectric- ferroelastic phase transition. As shown in the Section II.B.3, the measured splay elastic constant is indeed an effective elastic constant, smaller than the bare $K_1$ by the term $-\gamma^2\varepsilon_0\Delta\varepsilon_{low}$. For an increasing electric susceptibility, at a given temperature, the effective elastic constant will become zero and the uniform nematic phase will become unstable towards splay deformation. Such linear dependency between the effective elastic constant and the electric susceptibility is nicely reproduced by the experimental data (Fig. 6(b)).

### 4. Dielectric measurements

Dielectric measurements in the ferroelectric phase have to be taken with caution. Accurate measurements require successful aligning strategies, and as shown in section V, it is a challenging task in the case of the $N_F$ phase. For example, while in the N phase either some conventional homeotropic aligning agents or bare metallic electrodes seem to succeed in inducing homeotropic alignment, up to now, to our knowledge no homeotropic alignment was achieved in the ferroelectric state.

In either case, giant values ($\sim 10000$) of the dielectric permittivity have been reported for RM734 in the $N_F$ phase (see Fig. S5 in Ref. [68]). The situation can be compared with the dielectric measurements for DIO reported by H. Nishikawa et al. [15] and S. Brown et al. [73]. In this case, dielectric spectra in the N phase have been reported to be characterized by a single active relaxation process, whose amplitude, although large for standard nematics, increases only to values around 200 (see Fig. 4(e) in reference [73]). The transition to the $N_S$ (addressed as $N_x$ in [73]) phase results in the initial decrease in the amplitude, followed by a sharp diverging increase as the polar correlations grow. In the $N_S$ phase, an additional high frequency relaxation appears, which again disappears in the $N_F$ phase, which is characterized by a very strong relaxation process with a frequency of 2 kHz [73]. From the behavior of relaxation modes under bias fields, S. Brown et al. consider two scenarios. One in which local ferroelectric and antiferroelectric fluctuations coexist in the $N_S$ phase, or another in which the phase presents some polar order, but with an additional director modulation in the nm range, which results in the additional high frequency mode.

In either case, remarkably large dielectric permittivity values, of the order of 10000 have been reported in the $N_F$ phase for RM734, DIO, or comparable materials [15,68,73].

### 5. Polarization measurements

Such unprecedented values of the dielectric permittivity indicate the existence of large polar ordering in the $N_F$ phase. Polarization measurements have been performed via the standard triangular/square-wave method in which the field-induced polarization reversal current is measured. Despite the different approaches, i.e. use of in-plane or out-of-plane application of driving fields, reported polarization values saturate at remarkable values of the order of several $\mu C/cm^2$ : $6\ \mu C/cm^2$ for RM734 [20,75] (See Fig. 6(d)), $2.5 - 6\ \mu C/cm^2$ for DIO [15,73,75] or $6 - 7\ \mu C/cm^2$ (around $2\ \mu C/cm^2$ in case of planar cells with out-of-plane field) for an homologous of RM734 [83]. In RM734 and DIO mixtures, polarization saturation value varies continuously from one material to the other [75]. The reported polarization values are of the same order as those found for polar columnar phases [65], but an order of magnitude larger than common values found for classical ferroelectric phases [84] or bent-core polar phases [60].

### 6. Second Harmonic Generation

Further evidence of the strong polar character of $N_F$ is the extraordinary second harmonic generation (SHG) properties reported so far. Values of NLO coefficients ranging between 1 and 10 pm V$^{-1}$ have been reported for a wide number of materials from the RM734 (material 1a ) and DIO (material 5b) families, comparable to several ferroelectric solid materials (see comparison table S4 of Supporting Information of reference [68]). Recent SHG measurements of RM734 give a $d_{333}$ coefficient of 5.6 pmV$^{-1}$ [85]. As pointed out by C. Folcia et al. this value is one of the largest NLO coefficients reported for FLC despite the molecules not being specifically designed for NLO applications [85].

The temperature dependence of SHG intensity for DIO shows that in the absence of applied field, SHG intensity only appears in the $N_F$ phase (see Fig. 8). However, under the application of small electric fields, the $N_S$ phase becomes also SHG active [15]. The same trend is observed for materials investigated in [68] by J. Li et al. where SHG activity strongly increases in the $N_F$ phase in the absence of external fields. According to their results, the SHG signal increases over a wide range of temperatures around the transition, with a marked pretransitional behaviour.

The strong SHG activity has led to the implementation of SHG microscopy studies. RM734 N-$N_F$ transition has been studied via SHG microscopy (Fig. 2 in Ref. [19] and see Fig. 9 and Fig. SI.14 in Ref. [77] by N. Sebastián et al.), showing that on cooling from the N phase, the first detected SHG image corresponds to the stripe destabilization sequence at the transition described before, with a periodicity of few microns. Further cooling leads to the homogenization of the SHG texture, with a dependence on the direction of the incoming polarization well described by $cos^4\phi$, being $\phi$ the angle between the incoming polarization and the cell rubbing direction. Also, by means of SHG interferometry, it was determined that adjacent bands in the $N_F$ texture of RM734 correspond to opposite polarizations (Fig. S8 in Ref. [68]).

## 7. Direct isotropic to ferroelectric nematic transition

Notably, up to now, several materials have been found that exhibit a direct Iso-$N_F$ transition [66,71,73]. Materials showing such behavior span from monomeric compounds [66,73] to main-chain LC polymers [71]. In this case, the transition is characterized by a stronger enthalpy change (reported values for different materials are between 5.2 and 10 J/g) and exhibits some distinct characteristics.

Independently, A. Manabe et al. [66] and J. Li et al. [71] have shown that on cooling from the Iso phase, the transition into the $N_F$ phase occurs via nucleation of spherical domains (see Fig. 3 in [66] and Fig. 9 here). Such domains steadily grow on further cooling and eventually begin to coalesce and fill the whole area in a mosaic-like pattern made of domains divided by disclination lines, generally oriented along the rubbing direction. This process takes place over several degrees, e.g. 5 °C for material Nt4, picked as a representative example of the main-chain oligomers/polymers [71]. By means of SHG microscopy and interferometric microscopy, J. Li et al. show that in the circular domains nematic director forms a +1 defect with a concentric pattern, for which clockwise and anticlockwise directions of polarization can be found with equal probability. At lower temperatures, the mosaic texture is revealed to consist of polar domains with the alternating direction of polarization. Interestingly, longer oligomers/polymers show larger SHG intensity, varying between 50-times and 150-times the SHG intensity of Y-cut quartz.

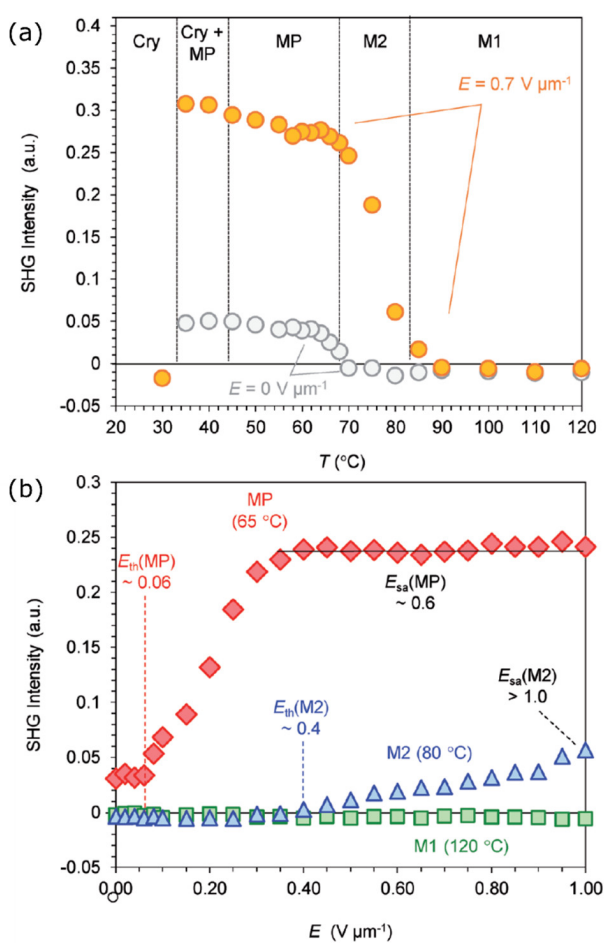

FIG.8. (a) Temperature dependence of the SHG signal of DIO for p-p polarization. (b) Field dependence of the SHG signal of DIO at three temperatures in the N (M1), $N_S$ (M2), and $N_F$ (MP) phases. © 2017 WILEY-VCH Verlag GmbH & Co. KGaA, Weinheim. Reprinted with permission, Ref. [15].

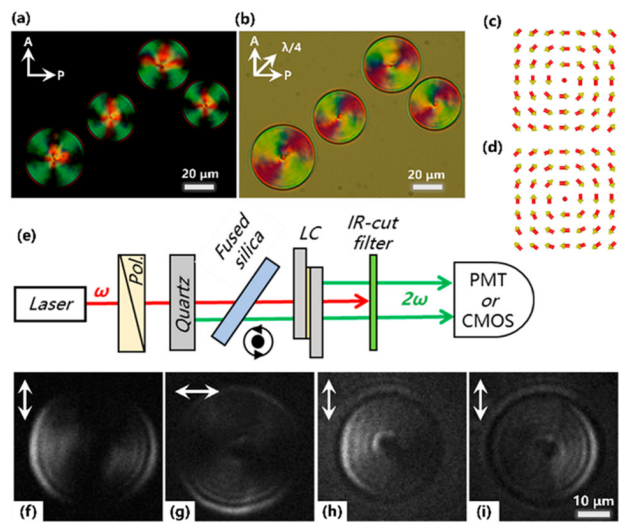

FIG.9. (a-b) POM images of domain nucleation on the direct Iso-$N_F$ transition. (c-d) Tangential anticlockwise and clockwise nematic director arrangement. (e) SHG and SHG-interferometry setups. (f-g) SHG microscopy images and (h-i) SHG-interferometry images for two complementary interferometry conditions. © 2021, American Chemical Society. Reprinted with permission, Ref. [71].

As before, the polar nature of the phase is also reflected in the dielectric properties. When cooling from the isotropic phase, dielectric permittivity grows from values around 100 in the isotropic phase to values up to for example 20000 in the case of a monomeric material [66] and in the range of 1000-10000 for the oligomers/polymers [71]. In the latter case, reported polarization values are in the range of $1.5\ \mu C/cm^2$, slightly lower than those for the monomeric materials.

### B. Chiral Phases

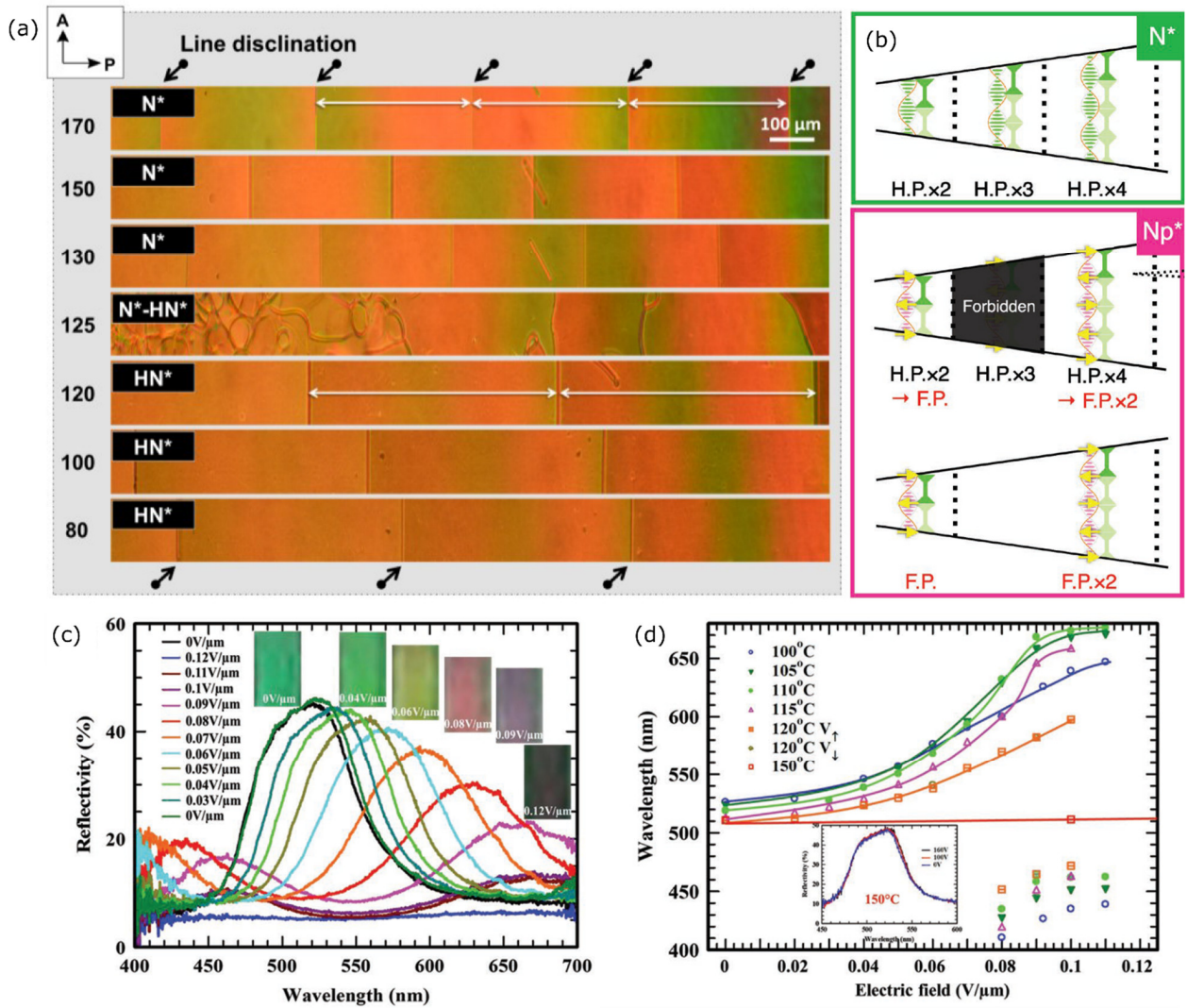

FIG.10. (a) Doubling of length between Gradjean-Cano lines in a wedge cell for RM734 doped with a RM734 chiral generator analogue. © 2021 Published under PNAS license. Reprinted with permission, Ref. [88]. (b) Schematic illustration of the proposed helical structures in wedge cells for N* and $N_F$* phases. © 2021 Wiley- VCH GmbH. Reprinted with permission, Ref. [86]. (c) Selective reflection for a RM734 + 2% BDH1281 mixture for different applied fields and (d) its voltage dependence at different temperatures. © 2021 Wiley- VCH GmbH. Reprinted with permission, Ref. [87]

Also remarkable, despite the early stages of the field, is the simultaneous realization of polar cholesteric phase $N_F^*$ [86–88]. Two different approaches have been described. H. Nishikawa et al. [86] and C. Feng et. al. [87], reported on the chiral ferroelectric nematic phase by doping DIO and RM734 respectively with commercially available dopants. Meanwhile, X. Zhao et. al. [88] doped RM734 with chiral generators, synthesized by introducing chiral groups at different positions of RM734. In either case, all the doped systems show a high temperature N* phase followed by the polar counterpart, in which the polarization as well as the nematic director twist along the helical axis (see Fig. 1).

In all the cases helical pitch smoothly varies with temperature, without drastic changes in the $N^*$-$N_F^*$ transition. Investigation of the pitch length via the conventional Grandjean-Cano method shows interesting behaviour of the Gradjean-Cano lines. In conventional N* materials, the distance between Grandjean steps (L) is determined by the helical pitch (p) by the relationship $L = p/(2tan\theta)$, where $\theta$ is the angle of the wedge cell. It should be noted here, that depending on the Burgers vector, L can be twice as large [89]. Nevertheless, it has been observed that at the $N^*$-$N_F^*$ transition, edge dislocation lines rearrange, and the distance between them doubles (see Fig. 10(a)). The underlying mechanism for such effect, far from being related to a drastic doubling of the helical pitch, has been attributed to the polarization direction at the surfaces. While in the N* phase half-pitch structures are allowed, in the polar $N_F^*$ even multiples of the half-pitch are needed to maintain the polarization in the same direction along the surface avoiding additional defects. Thus, in the $N_F^*$ it is proposed that the following relationship is valid $L = p/(tan\theta)$ [86,88] (Fig. 10(b)). Additionally, the coupling between the helical structure and the polarization results in enhanced control of the tunability of selective reflection of visible light. Fields as low as 0.02-0.1V/μm are reported to be enough for reversible tuning [87] (see Fig. 10(c)). Also enhanced performance is reported in terms of switching times for optical transmittance, with 10%-90% On and Off times of the order of 250 μs respectively, with a reduction for the off time to 15 μs in the case of using a reverse pulse field [86].

To conclude this section, it should be mentioned that in addition to the described phase sequences, two additional polar phases have been reported for an RM734 analogue, with the addition of a fluorine group in the center benzene ring by R. Saha et al. [83], following the phase sequence on cooling Iso-N-$N_F$-$F_2$-$F_3$. Such phase sequence was determined via DSC measurements, combined with spontaneous polarization measurements. Although the material is the same as compound 2 reported in [73] and the difference could arise from different purities, such results highlight the possibility of discovering additional polar phases in the present context.

### C. MD simulations

Computer simulations play an important role in the understanding of the formation of liquid crystalline phases and their properties [90,91]. Immense progress in computer power and advanced force fields models enabled all atomistic molecular dynamic simulations of various LC phases. As the formation of orientational order in a liquid is a many-body problem, which is difficult to address with the theory, the molecular dynamic (MD) simulations give insight into dynamics and correlations between the molecules which are a consequence of many-body interactions.

The material RM734 was studied by two independent MD simulations by X. Chen et al. [20] and R. Mandle et al. [70]. The simulations differ in choice of the force fields (in Ref. [20] APPLE&P [92,93], in Ref. [70] GAFF-LCFF [94,95]), the number of molecules per simulation (368 in Ref [20] versus 680 in Ref. [70]), and production MD simulation length (about 20 ns in Ref [20] versus 250 ns Ref. [70]). Both simulations failed to observe the transition between the apolar nematic to polar nematic state, so the simulation of either polar or apolar state was determined by starting configuration. Both simulations of the polar state of RM734 gave comparable values of the order parameters $S$, $<P_1>$ and the spontaneous polarization, and similar positional correlations showing stronger polar head-to-tail pairing in the polar than in the apolar simulation (Fig. 11). The position of the X-ray scattering peaks calculated from the polar MD simulation by R. Mandle et al. matched well the positions of the measured low angle peaks (Fig. 6(c) and Fig. 6 in Ref. [70]), while those from apolar did not [70], which supports the relevance of MD simulations. As the low angle peaks are experimentally observed also in the nonpolar nematic phase well above the phase transition temperature to the polar phase, this indicates that strong polar correlations are present already there. These observations suggest that small angle X-ray scattering could be a useful probe of polar order within apolar and polar NLC.

The differences between the positional correlations in the polar and apolar states, on their own, don't explain why in a particular material the polar state is more favorable than the apolar. However, in the simulations in Ref. [70], a notable difference between the states was observed in the density and mobility of the molecules. In those simulations, the density was about 0.5 % larger, and the diffusion of the molecules was faster in the polar than in the apolar state (Fig. 11(d)). The increase of the density was also confirmed experimentally for RM734 (Fig. 6(f)), as well as for a fluorinated homologue [83]. On contrary, in the material RM734-CN, which is very similar to RM734 but doesn't exhibit a polar phase, the simulation showed that the polar and apolar states have the same density, while the diffusion is slower in the polar state. This indicates that in the materials exhibiting the polar phase such as RM734, the molecules can pack better and, simultaneously, they are more mobile when in the polar state, which suggests that the polar phase is more favorable than the apolar from the point of view of the excluded volume. These observations were confirmed by simulations of several other materials from the RM734 family [70] and also for DIO [96].

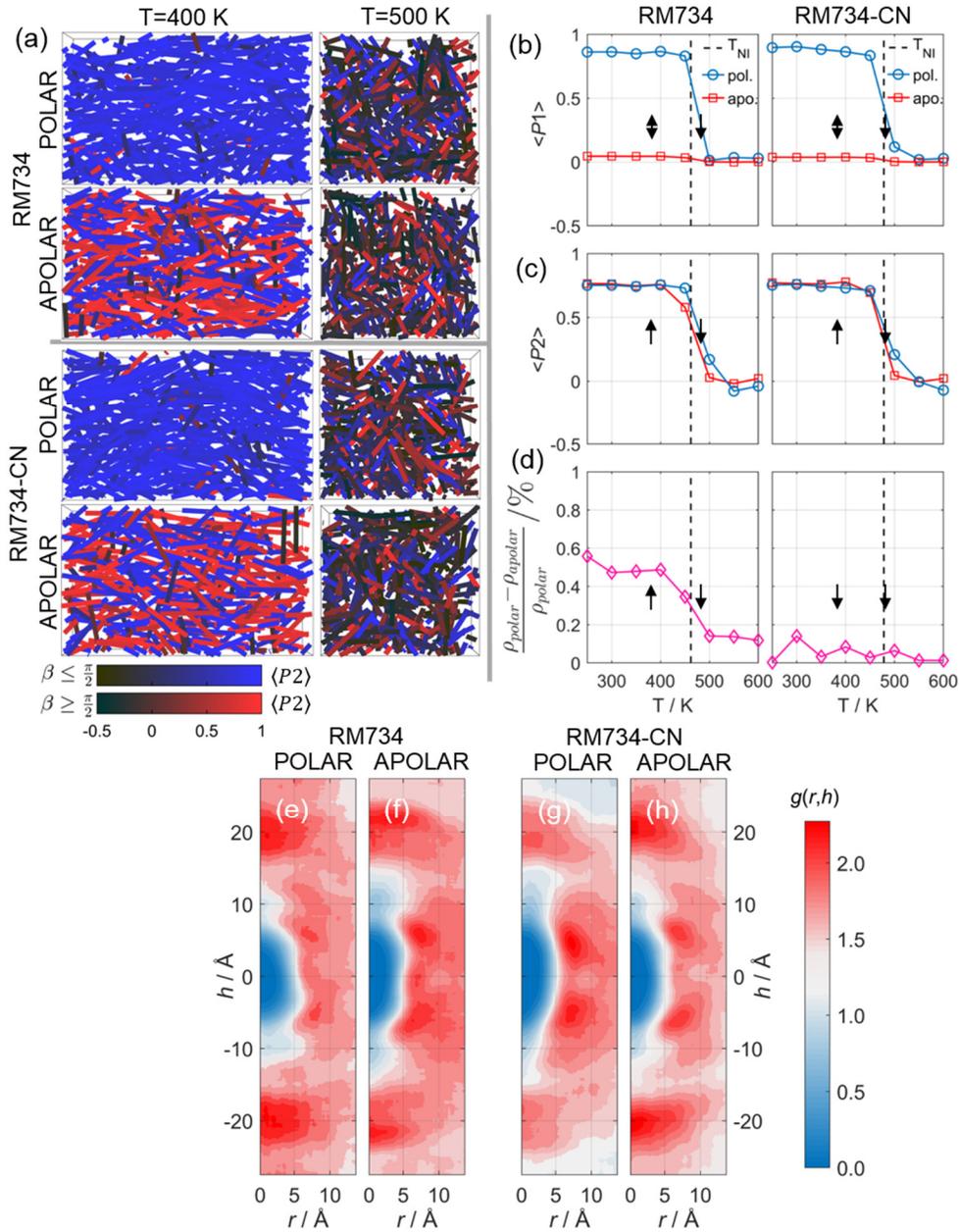

FIG.11. a) Molecular dynamic simulations snapshots in polar and apolar configurations of RM734 and RM734-CN at two temperatures, 400 K and 500K. Dependence of the order parameters P1 (b) and P2 (c) on the simulation temperature for both materials in both, polar and apolar configurations. d) Difference in the simulated mean density between polar and apolar configurations. e-h) Cylindrical pair-correlation functions as obtained for both materials and polar and apolar configurations at 400 K. Reproduced from Ref [70] licensed under CC BY 4.0.

## V. ANCHORING AND ELECTROOPTICS

The switching behavior of any LC is strongly affected by the confining surfaces. In the case of ferroelectric nematic phases, the interaction with the surfaces is additionally complex due to electrostatic effects.

### A. Surfaces and anchoring effects

The importance of the confining boundaries was systematically studied by Caimi et. al. [80], where the alignment and structures induced by different alignment surfaces were explored and surfaces assessed by their degree of coupling strength and anchoring orientation. It is evidenced that the textures and defect structures, together with the field response are strongly dependent on the surface treatment.

While silane-coated surfaces are usually employed in thermotropic nematics to achieve homeotropic alignment, they induce planar alignment in the case of the N phase of RM734, homogeneous when rubbed [80]. In the $N_F$ phase, both rubbed silanized and rubbed Teflon surfaces, show strong quadrupolar orientational coupling with the RM734 molecules, and to a less extent polar coupling (with the direction of polarity antiparallel to the rubbing direction), resulting in homogeneous areas of planar alignment in which embedded twisted domains can be observed. The latter correspond to areas in which one of the surface polarity is reversed. From the different aligning strategies, only bare hydroxylated glass resulted in homeotropic alignment in the N phase, which turns into inhomogeneous planar alignment in the $N_F$ phase.

Homeotropic alignment in the N phase was also reported for DIO and RM734 for two different cells: treated with octadecyltrimethoxysilane from EHC Japan [15] and with bare ITO electrodes [19]. In both cases, transition to the $N_S$ or $N_F$ phase was accompanied by the development of a sandy/grainy texture with no clear extinction position [15,19].

Commercially available LC cells for planar alignment most commonly employ rubbed polyimide layers with a small pretilt, necessary for defect-free alignment. While pretilt results in equivalent surface energies for **n** and –**n**, it also results in non-zero polar surface energy, differentiating between **P** and –**P** states as pointed out by X. Chen et al. [78]. Cells can be assembled with surfaces having parallel or antiparallel rubbing directions, which has been shown to result in different textural characteristics. In either case, strong planar alignment is observed in the N phase. In the $N_F$ phase, parallel rubbing would orient the polarization in the same direction on both surfaces, resulting in a monodomain uniformly aligned sample (see Fig. 2(f-i) in reference [78]). On the other hand, in antiparallel rubbed cells, in-plane polar orientation at both surfaces with opposite directions is expected. When cooling from the N phase (see description of the transition in previous sections), initially the $N_F$ phase is characterized by a homogeneous texture, appearing right after the stripe destabilization (see supplementary videos 1 and 2). At a lower temperature (very close to the transition in the case of RM734 and a few degrees after in the case of DIO) a wall front propagates through the cell, usually initiating at some surface defects. When two of the fronts come together, they can either annihilate or give rise to a $2\pi$ wall dividing domains with opposite twist handedness [77,78]. When those walls are perpendicular to the rubbing direction, they stretch over long distances undeformed. However, when directed along the rubbing direction they adopt a sierra-shape configuration, with the wall bending at angles between 50 and 70 degrees as shown by N. Sebastián et al. [77]. Uncrossing of the polarizers reveals opposite optical behavior in both domains, characteristic of twisted domains with opposite handedness [73,77,78]. In such domains, polarization rotates a $\pi$ turn, in either, left- or right-handedness direction. The simplest of the twist structures is a linear $\pi$-twist across the cell thickness, although the symmetry of the dependence of the relaxation rates of the director fluctuations on the wave vector, reflecting the symmetry of the structure, points towards more complicated structures [77]. Nevertheless, these twisted domains can span up to several mm, and a single domain can even extend over a whole cell on some occasions.

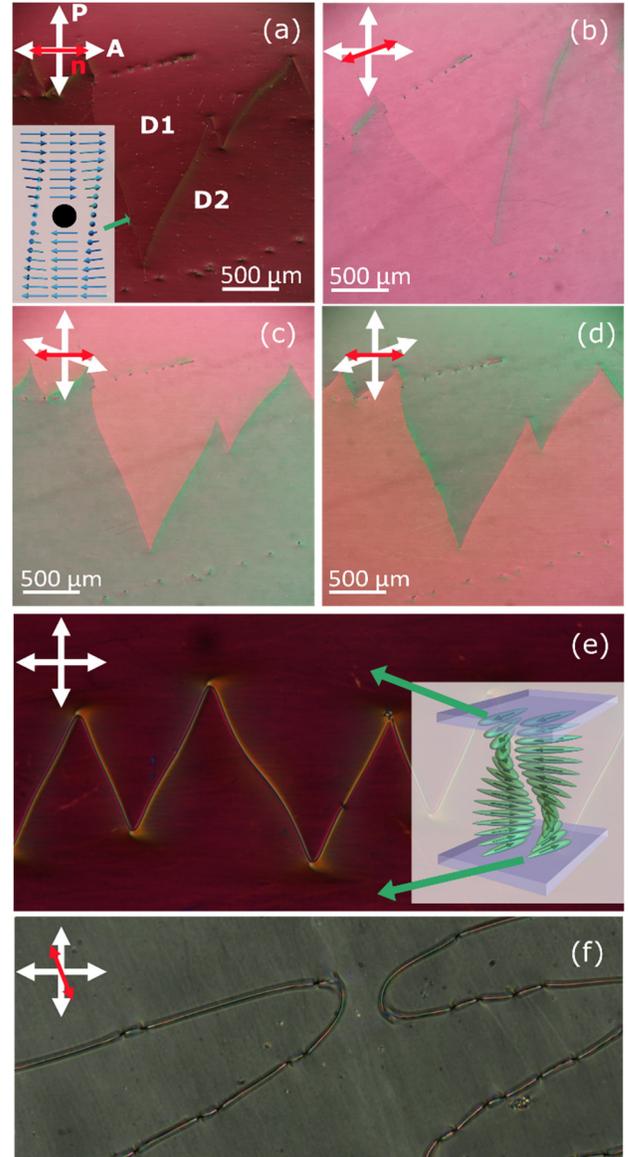

FIG.12. (a-d) $\pi$-twist domains of opposite handedness in the $N_F$ phase of RM734 in extinction position, with the sample rotated and on uncrossing analyzer clock- and anticlockwise. (e) Detail of sierra-shape wall. (a-e) Reproduced from [77], published under CC BY-NC-ND 4.0). (f) Point-like structures decorating walls. Reproduced from [18], licensed under CC BY 4.0.

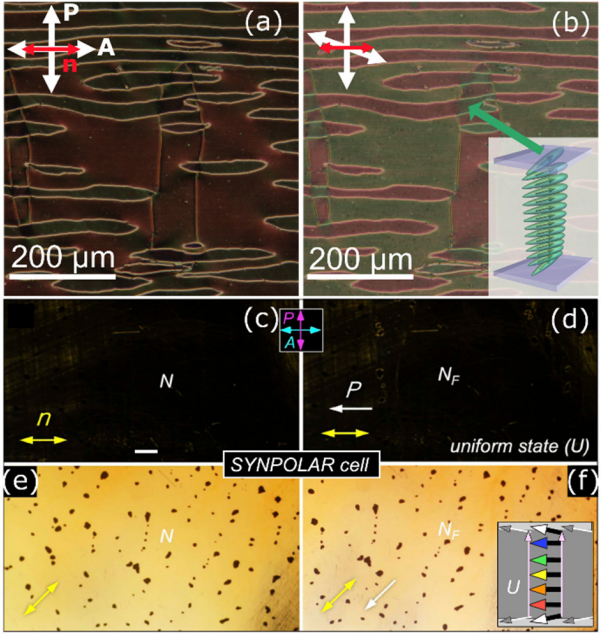

FIG.13. (a-b) Simple twist structures surrounded by domain walls pinned at the surface. Reproduced from [77], published under CC BY-NC-ND 4.0. (c-f) Textures of the N and $N_F$ phase in a SYNPOLAR cell (parallel rubbed cells). Reproduced from [78], published under CC BY-NC-ND 4.0.

Additionally, textures in cells with surfaces characterized by bidirectional rubbing have also been studied by X. Chen et al. [78]. Bidirectional rubbing results in non pretilted surfaces, with a preferential parallel orientation of **n** along the rubbing but no in-plane polarity. In the $N_F$ phase, both directions of polarization are then equally probable and the texture is characterized by domains of opposite polarization direction [78].

### B. Electro-optic behaviour

Electro-optics in liquid crystals is of utmost relevance, as it constitutes the basis of their technological potential. In the polar nematic phases, unique features are to be expected. However, is to be noted that electro-optic behavior will strongly depend on the confining boundaries determining the corresponding director structure and on the direction of field application.

Considering out of plane fields applied in planarly aligned cells, it has been shown that, due to its large value, reorientation and deformation of polarization results in an induced charge that neutralizes the applied field hindering the reorientation (see Section II.B.4-Electrostatics). This results in very large voltages needed for splay-bend reorientation of the director when compared with the standard N phase (see Section 6, Supporting Information in reference [20]). However, interesting behavior can still be observed at lower voltages [77]. In antiparallel planar rubbed cells, where cells are covered by $\pi$-twist polar domains, the application of smaller fields (low frequency AC fields of amplitude $\sim 0.3 V/\mu m$) causes the deformation of the domain wall, enhancing the sierra-shape configuration of those running parallel to the rubbing direction or triggering the growth of new line defects, that eventually open and result in new domains. For applied voltages oscillating between 0 and V, it can be seen how domains of one handedness progressively grow, while the opposite domains shrink. Reversal of field polarity (-V to 0), consistently reverses the process.

The richness of polarization-director structures is also reflected in the electrooptic behavior upon application of in-plane fields. In the predominant $\pi$-twist domains, X. Chen et al have described three distinct electrooptic modes [78]. The application of an electric field perpendicular to the rubbing direction in the surfaces (i.e. the field is perpendicular to the director in the surfaces but parallel to **n** in the center of the cell), reorients the molecules, squeezes the twist to the cell surfaces [77,78] and gives rise to a quasiuniform state. Such reorientation is shown to be fast, with a characteristic time scaling as $\tau = \gamma/PE$, resulting in around 100 microseconds for fields of $\sim 10\ mV/\mu m$ [78] (Fig. 14(a)). In commercial IPS cells, with interdigitated electrodes, such mode results in opposite switching characteristics for adjacent electrode gaps and mirrored across the domain walls dividing domains of opposite handedness [77] (Fig. 14(b)). While short pulses minimize the effect of mobile charges, application of slower alternating fields shows bimodal response, as ionic impurities have time to redistribute, screening the field and the polarization charges.

If the in-plane field is applied opposite to the polarization in the quasiuniform state induced by the field in the previous case, the response starts in the twisted areas close to the surface and propagates through the cell as solitons in the polarization field until creating a twisted region in the center of the cell [78]. Such reorientation is much slower than the previous molecular reorientation and scales with the applied field also as $\tau \propto 1/E$ [78]. Additionally a third mode has been described, similarly to the observations for out of plane fields, consisting of the nucleation of domains with more favorable twist handedness depending on the field polarity (Fig. 14(c) reproduced from Fig. 4 in reference [78]) and their propagation throughout the cell. The monodomain is maintained after the field removal resulting in bistable switching between domains of opposite handedness.

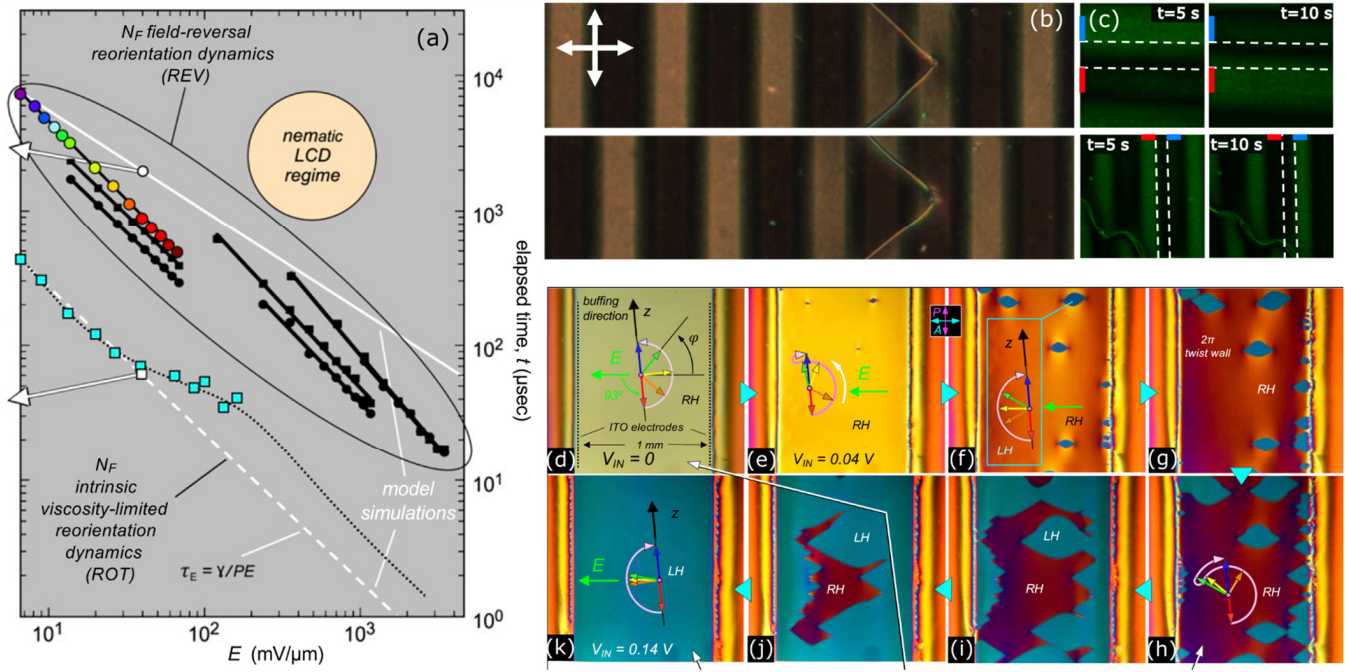

FIG.14. (a) Experimental and simulated reorientation times for the polarization reversal and viscosity-limited electro-optic modes. Reproduced from [78], published under CC BY-NC-ND 4.0. (b-c) Electrooptic response (POM and SHG) of two adjacent $\pi$-twist domains in an IPS cell with rubbing direction parallel to electrodes. Reproduced from [77], published under CC BY-NC-ND 4.0. (d-h) Domain-mediated polarization reversal of RM734. Reproduced from [78], published under CC BY-NC-ND 4.0.

## VI. EXPERIMENTAL CHALLENGES

In this section, several experimental techniques relevant to the present field are introduced. The different challenging aspects of such measurements are outlined together with the consideration of currently absent studies.

### A. Materials

One of the first steps for the development of the field relies on the ability to develop materials exhibiting the ferroelectric nematic phase. As evidenced by the time that has passed from the prediction to the experimental realization, this is indeed not a straightforward task. Systematic studies have shown that small molecular variations can suppress the $N_F$ phase [68,69]. For example, R. Mandle et al. have shown that a large dipole moment or polarizability is not sufficient for the appearance of the polar phase, and the removal of a single ester unit or lateral chain implies the disappearance of the phase [69]. The strong potential for the applicability of polar nematic phases in technological solutions relies on the ability to synthesize new materials exhibiting the ferroelectric phase at broad temperature ranges around room temperature. Few first examples are close to it [66,71] and constitute a very promising starting point. There is a compelling need to extend and deepen the knowledge about which fundamental features of molecular design are required for the promotion of polar nematic phases. As further discussed below, molecular dynamic simulations might have an important role in this task. Should be noted here, that during the writing of this paper new materials have been reported [97] by exhibiting ferroelectric nematic phases, evidencing the velocity at which the field is evolving. Promising progress has been done in recent years and we believe, that the field will advance fast with the development of new materials and probably more importantly, mixtures with improved applicability.

### B. Polarizing optical microscopy and anchoring effects

One standard method for the imaging of optically anisotropic materials is polarizing optical microscopy (POM), in which the sample is placed in the optical pathway between polarizers. In the standard configuration, the polarizers are kept orthogonal. When propagating through the sample, the incident linearly polarized light splits into two perpendicular components, travelling within the sample at different velocities. The properties of elliptically polarized transmitted light are determined by the second polarizer – analyzer. In the case of liquid crystals, this analysis provides information on the underlying director structure [98].

For this purpose, liquid crystals are typically studied under confinement in thin cells, in which the boundary conditions (for example chemical, steric or dipolar interactions) at the surface determine the orientation of the molecules at the interface, which for thin samples, is kept via elastic forces across the cell [99]. As covered in Section V.A, surface and anchoring effects in the $N_F$ phase are not only determined by the orientational coupling in the surface, but also by the polar coupling. This leads to different orientations and confined structures than in the classical nematic phase for standard aligning techniques. A strong tendency for the appearance of twisted structures is also evidenced. Obtaining a detailed analysis of the underlying structure from the interpretation of POM images of inhomogeneous birefringent media is complicated. Provided the structure is known or can be modeled, one can always rely on excellent numerical tools that have been developed to calculate POM images of arbitrary director structures [100,101]. That is, it requires the knowledge of the structure to simulate POM images that will match the experiment, and does not work vice versa in complex cases, where the exact structure cannot be deduced from the POM experiments. This is especially the case, when complicated twisted structures are involved, as POM experiments only give integrated information across the cell thickness. And thus, the development of an adequate theory for the $N_F$ phase (Section VII.C) is of utmost importance.

LC phases typically exhibit distinct optical textures such as domains and defect structures that enable their differentiation using POM. Distinctive anchoring of the $N_F$ phase offers direct ways of its identification. In Fig. 12, typical defect structures observed in the $N_F$ are shown. The most characteristic are the "sierra shaped" walls that appear in thin antiparallel rubbed cells and separate the π-twisted domains with opposite handedness. In thicker cells (depending on the material, typically > 10 μm), these walls are more curved and are typically decorated by point-like structures. They are also less stable, they tend to annihilate and often, after the relaxation, a whole cell consists of a single π-domain. In a thick cell, such domains may show good extinction between crossed polarizers because of the guiding of optical polarization, so one should be careful not to interpret such domains as uniform. An additional method for identification of the $N_F$ phase via microscopy observations has been proposed by P. Rudquist [102] using a hybrid planar cell, with a uniformly rubbed surface on the bottom and a circularly rubbed pattern on the top one.

One should also note here that strong memory effects can be observed in the cells reported in Section V after several heating and cooling runs, strongly affecting the resulting textures. The large polarity of the $N_F$ phase could be held responsible for such effect, rewriting the initial polarity of the aligning layer. Additionally, although uniformly aligned samples can be obtained in some cases, appearance of unintentional surface imperfections can be triggered by surface imperfections. This brings us to one of the important challenges to be faced. Currently, a bottleneck for the characterization of the phase and its exploitation is our ability to control the alignment and structures in the $N_F$ phase. This applies for simple uniform planar or homeotropic structures, which are critical for the determination of some material parameters such as NLO coefficients, dielectric constants or polarization values. Or for more complicated structures, for which precise control of the orientation of the director would also lead to the control of the orientation of the polarization, and thus tailored polarization structures. Compared to solid ferroelectrics, polarization, in this case, can vary continuously in space.

### C. Second Harmonic Generation

Optical studies of the response of the material to an electric field can infer ferroelectric order but their drawback is that assumptions about the structure and its field response have to be made. So they alone cannot be used to unambiguously prove that the phase is ferroelectric. For that, SHG studies and/or measurements of the hysteresis loop are needed [103]. In this respect, SHG is an essential tool for assessing the structural polarity of materials, as only those lacking the center of inversion symmetry allow SHG to appear.

SHG is a nonlinear optical process, in which a photon with frequency $2\omega$ is generated in a material via the absorption of two photons of frequency w, i.e. frequency doubled light emerges from a material pumped by a laser beam. Such a process is coherent, and in the case of molecular systems, it is determined by the orientational average of the molecular third rank hyperpolarizability tensor. In comparison to electric measurements, the advantage of SHG is that it is a noninvasive optical method, which detects polar order also in the absence of an external field and, therefore, ionic impurities do not affect the measurements. By using SHG interferometry option, the method is also sensitive to the sign of **P** (eg. see Fig. 9). Combining the investigations of field responses with SHG measurements gives also information on induced polarization (eg. see Fig. 8), which is in electrical measurements often obscured by ion contributions.

Combined SHG microscopy and interferometry are called to be essential and of great interest in the case of polar nematic phases. Not only it will bring valuable information for the development of the theory on electric polarization orientation but also for future design and exploitation of the materials' SHG capabilities. As mentioned above, the calculation of light propagation in non-homogenous birefringent media is complex and requires numerical methods. The analysis and calculation of the generation and propagation of SHG light in such structures get additionally difficult, as coherent effects need to be considered. Currently, there are no appropriate numerical tools for such studies, and they need to be developed.

Such progress will also be remarkably useful in terms of electrooptic behavior studies. The electrooptic behavior described in Section V.B is expected to be more broadly and deeply analyzed with the development of new materials. The large polarization values result not only in complex linear electrooptic behavior but also entail the corresponding second harmonic electrooptic switching as demonstrated in reference [77] (Fig. 14(b)). This is expected to be the subject of many future studies.

### D. Dielectric Characterization

Broadband dielectric spectroscopy is an exceptionally useful experimental technique for the investigation of liquid crystal phases and phase transitions, in which measurement of non-collective and collective orientational polarization contributions to the dielectric spectra allows to study the dynamics of the material [104]. Non-polar nematic phases are characterized by two molecular modes of orientational polarization, i.e. molecular rotation around the long and short molecular axes, contributing differently to the parallel and perpendicular components of the macroscopic dielectric permittivity. In the present case, broadband dielectric spectroscopy measurements through the non-polar nematic and different polar nematic phases are of utmost interest. As has been shown in Section III, the onset of collective behavior can be well found in the non-polar nematic phase (whose dielectric behavior greatly differs from that of "classical" nematics). Detailed studies and comparison between different materials will allow studying the development of spontaneous polarization through the phase transitions, and should shed light, into which molecular correlations are determinant for the development of one or another phase sequence. Such comparison together with studies of thickness dependence effects are currently missing but needed.

Proper characterization of the parallel and perpendicular components requires successful aligning strategies, which as discussed above remains a challenge. In this context is important to remark that, as pointed out by Brown et al. [73], while the effect of the thin aligning layers of the measuring cells in the measured values can be disregarded for conventional liquid crystals, that is not the case for the present ferroelectric materials and can lead to considerably different values. Additionally, the very small values of the splay elastic constant result in considerably low threshold voltages for Fredericks transition in the high temperatures N phase already far from the transition. On lowering the temperature, $V_{th}$ is further decreased. Thus, typical probe voltages (0.1-0.5 $V_{rms}$) can fall above $V_{th}$ leading to field induced changes in orientation during the measurements. Consequently is important to pay attention to this detail, taking into account that adequate reduced voltages of around 20 $mV_{rms}$ can also lead to noisy signals. For the appropriate understanding of the molecular polar correlations, it is necessary the careful measurement of the broadband spectra, its analysis and the identification of the relaxation modes. Finally, such careful analysis and interpretation should identify the modes associated with the reorientation of large polarization values that can lead to misinterpretation of epsilon values [105]. Due to all of this, to the difficulties in achieving homeotropic alignments, the different geometries or substrates used by different research groups and to the presence of free charges, the interpretation of the absolute permittivity values and comparisons between materials should be considered very carefully.

### E. Polarization measurements

For the determination of spontaneous polarization ($P_S$) in liquid crystalline samples, the triangular (square)-wave method is broadly used, in which $P_S$ is determined from the analysis of transient currents under triangular(square) AC voltages [106]. Determination of $P_S$ requires of the accurate deconvolution of the different conductive, capacitive and polarization terms. While triangular fields result in a linear capacitive current contribution, such contribution in the case of square-voltage is often more complicated than a single exponential, which could lead to overestimation of the polarization value. Additionally, there are several limitations to the technique. Election of wave amplitude and frequency is also determinant in these experiments, where the measuring frequency should be lower than the inverse of the polarization relaxation time, but large enough to avoid effects produced by charges. Application of large voltages can lead in this case to turbulent flow, which will also contribute to the transient currents. The high polarity of the sample, additionally imposes some limitations. As pointed out in reference [83] difference between polarization values obtained for in-plane and out of plane cell geometries (See Section IV.5) is a consequence of the large internal electric field that needs to be surpassed in order to rotate the polarization out of plane, which can lead to underestimated values for out of plane geometries. On the other hand, non-uniformity of the field in the case of in-plane cells, for which the electrode does not extend all the thickness, can lead to non-accurate determination of the area, leading to non-accurate values of the polarization. Because samples with homogeneously aligned polarization in the absence of the field are very difficult to achieve, determination of the value of the spontaneous polarization, that is polarization in the absence of the field, is challenging. In particular, this poses a problem in the temperature interval where the polarization value changes significantly and the polarization value is a nonlinear function of the applied field.

## VII. FUNDAMENTAL CHALLENGES AND FUTURE DEVELOPMENTS

In this section, several important fundamental open questions are presented. One of the very important questions that should be addressed in future years is the mechanism by which polar ordering originates in these phases. At this respect, some thoughts and contributions are collected here for consideration. Strong pretransitional behavior as discussed above carries information of the physical mechanisms that drive the phase transitions. Interpretations of the differences in the pretranstional behavior, which lead to either the $N_F$ or $N_S$ phase, is a challenge commented first. It is followed by some remarks on currently observed phase sequences, pointing out the challenge of demonstrating the uniformity of the polar ground state, fundamental for the adequate identification of the polar phases. There is also place for a short discussion about phase nomenclature, regarded as necessary at this stage. Also of utmost importance is that the understanding of these phases, description of their properties and prediction of structures, defects or field responses requires of development of a suitable theoretical model, currently not available. This is a critical challenge, which is here just outlined with some aspects that should be important to consider. Development of the experimental aspects of this new field will be strongly intertwined with the development of a suitable theory. Finally, success in both, experimental and fundamental understanding of these new materials will determine their future applicability. Here, as a short recap, we have collected several thoughts on applicability of polar nematic phases into technological solutions based on some contributions that outline remarkable exploitable properties.

### A. Pretransitional behavior: the role of the splay in the phase transitions

#### 1. Pretransitional behavior

The transition from N to $N_F$ phase has been shown to be a ferroelectric-ferroelastic transition in which the growth of ferroelectric order is accompanied by the softening of the splay orientational elastic constant [18,19]. The orientational fluctuations of **n** are characteristics of the nematic phase and, because the NLCs are optically uniaxial with the optical axis parallel to **n**, these fluctuations manifest optically as distinctive nematic flickering. In a bulk NLC, their eigenmodes are overdamped plane waves with two branches: splay-bend and twist-bend [38]. Relevant to our discussion are the pure splay modes schematically shown in (Fig. 15(e-g)). If the shape of the constituents lacks head-tail symmetry, the splay fluctuations cause that locally more constituents are oriented in the same direction (Fig. 15(f)). If they also carry longitudinal dipole moment, this ordering will result in local electric polarization. This is a well known flexoelectric effect [44]. On the other hand, for some reason, e.g. steric and/or electrostatic interaction, the polar order of several neighboring molecules may become favorable, which manifests as a collective mode observed in dielectric spectra. This mode can be described as fluctuations in the magnitude of the polarization (Fig. 15(b-c)). By flexoelectric coupling, required by symmetry, this mode is coupled to splay fluctuations, so instead of independent polarization and splay mode, there are two coupled modes. One is primarily the splay mode, seen optically, the other is primarily the polarization mode observed by dielectric spectroscopy. Because of the coupling, splay fluctuations promote the growth of the polarization (Fig. 15(e)), and vice versa, polarization fluctuations promote splay, which results in softening of the effective splay elastic constant. This seems to be the driving for the transition from the nematic to the polar nematic state. It also explains why in MD simulations, the transition between nematic and polar nematic is not observed because due to small size and periodic boundary conditions, the splay fluctuations are prevented. As shown in Section II.B.3, this mechanism predicts the appearance of the 1D or 2D modulated antiferroelectric splay nematic phase $N_S$ as an intermediate phase between N and $N_F$ [18,19,28]. The models that use the lowest order coupling terms in the free energy (Eqs. 8 and 9) predict the second order phase transition with the modulation wavevector growing from 0 at the N-N$_S$ phase transition. By adding higher order coupling (gradient) terms, the N-N$_S$ transitions can become (weakly) first order with the finite modulation wavevector at the phase transition.

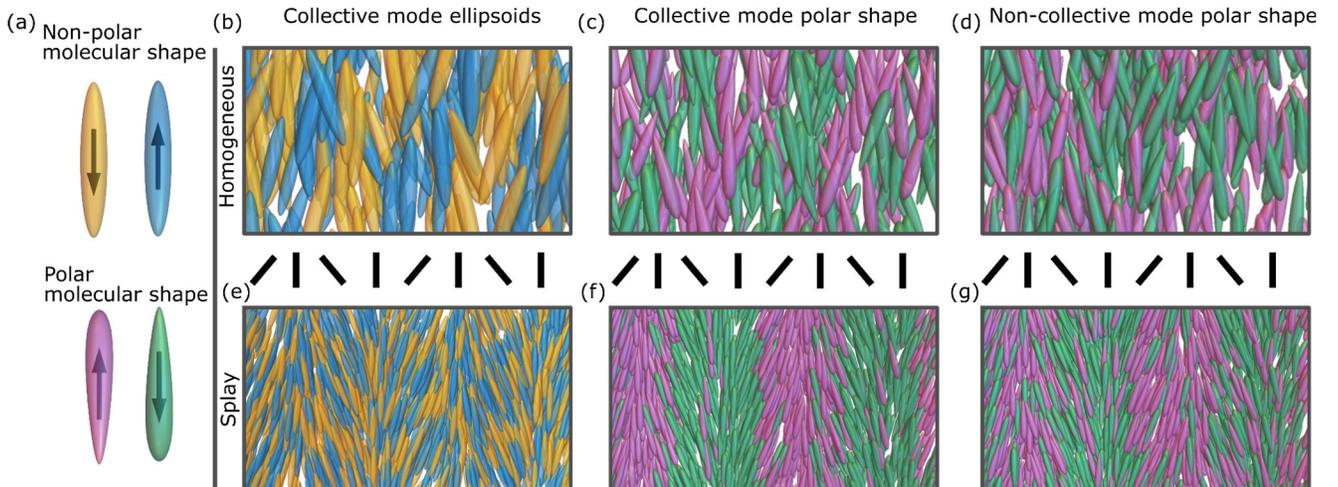

FIG.15. Fluctuation modes (a) Color code for different up-down orientations of molecular constituents having head-tail symmetry (blue-yellow ellipsoids) and lacking head-tail symmetry, i.e. polar-shaped (wedge shaped pink-green). Collective mode for (b) non-polar shaped and (c) polar-shaped molecules. (d) Without collective mode (no polar correlations) for polar-shaped molecules. Splay fluctuations of (e) non-polar-shaped molecules showing no coupling of splay and polarization, while the flexoelectric polarization is (f) enhanced in the presence compared to (g) the absence of collective polarization mode.

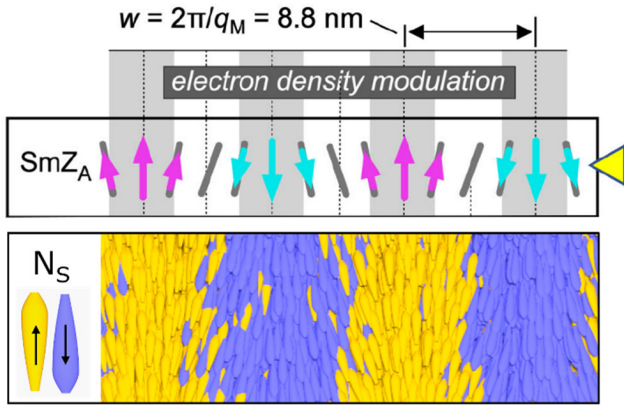

Fig. 16. (top) Schematic representation of the proposed SmZ$_A$ structure, where grey shadowed areas have a different density than the layer boundaries (white shadowed areas). Reproduced from [107], licensed under CC BY 4.0 (bottom) Schematic representation of the splay nematic phase N$_S$.

The study and comparison of pretransitional behavior for materials with different phase sequences or temperature ranges of intermediate N$_S$ phase should lead to discern the mechanisms responsible for a given phase sequence and the origins of the phase polarity.

### 2. Phase sequences

As will be discussed in the paragraph below, in the case of DIO, the phase, which appears in between N and N$_F$, has indeed been identified by X. Chen et al. as modulated splayed phase [107]. This leaves us with the question, of whether the phases in RM734 also follows the same sequence. In our first publications, the transition was identified as N-N$_S$, guided by the unambiguous role of the $K_1$ softening in the transition and appearance of a stripped polar structure observable by SHG microscopy in a narrow temperature region of a few tenths of a degree before undergoing the transition to what is now known as N$_F$ phase. Recent precision adiabatic scanning calorimetry measurements by J. Thoen et al. [74], show a very weakly first order transition, where pretransitional variations are significantly larger in the higher temperature phase. Optically, the transition from N to N$_S$ phase would be characterized by the freezing of the splay fluctuations. Optical observations under slow cooling (See Supplementary Video 2) show for RM734 a narrow but discernible temperature range in which splay fluctuations are frozen, in between the N phase and the low temperature phase. Such texture is compatible with the N$_S$ phase observed in DIO. This raises the question of whether the transition is a direct N-N$_F$ transition or N- N$_S$ -N$_F$ double transition, maybe in the case of the confined sample, stabilized due to surface effects. Additionally, ionic impurities could play an important role in the stabilization of a narrow N$_S$ phase. This will be, for sure, the focus of future research.

On the other hand, the question of whether the experimentally observed N$_F$ phase is indeed uniform remains open. In comparison to the nonpolar nematic phase, in which the uniform state is easily achieved in planar or homeotropic cells, this is not the case in the N$_F$ phase. There under confinement and different aligning strategies, the director structure shows a strong tendency to twist and to form topological line/wall defects. These may be the consequences of surface and electrostatic effects. However, one should bear in mind, that in the N$_F$ phase, due to the symmetry, in the elastic free energy linear terms in deformation are allowed which might destabilize the uniform alignment [7]. In this regard, proving the uniformity of the ground state is challenging as experimentally accessible N$_F$ structures will always be affected by surface and electrostatic effects.

Finally, given that we are currently at the early stages of the research in ferroelectric nematic materials, it can not be ruled out that the progressive development of new materials, from monomers to polymers, will lead to the discovery of additional phases.

### 3. Intermediate phase and nomenclature

In a recent preprint [107], the intermediate phase in DIO has been shown to exhibit periodic modulation of density in the direction perpendicular to **n** with the periodicity of 8.8 nm. X. Chen et al. propose the structure of the phase being layered antiferroelectric, with the nematic director and polarization oriented parallel to layer planes, and the polarization alternating in sign from layer to layer accompanied by alternating splay (Fig. 16, top). They named the phase SmZ$_A$. The proposed structure is identical to the splay nematic phase N$_S$ (Fig. 16, bottom), although the authors in Ref [107] did not make that connection. Taking into account that the density depends on the degree of polar order as demonstrated by R. Mandle et al. via MD simulations [70], it is expected that the N$_S$ phase exhibits density modulation. No doubt, the naming of the phase will be a subject of future debates. In our opinion, there are two arguments to take into account. First, the N$_S$ phase, as shown in section II.B, in its origin belongs to the family of modulated nematic phases such as the N$_{TB}$ and N$_{SB}$ phases. And it is important to note that density modulation is also expected for the N$_{SB}$. Modulated N$_S$ phase as schematically shown in Fig. 16 is predicted by the theory by several authors [18,28,51] where periodicity of the modulation depends on the material paramters. Additionally, the intermediate phase in DIO appears to have a modulation period about 40 times larger than the molecular width, while thermotropic smectic phases typically exhibit a modulation period of the order of the molecular size. With this in mind, we believe that to avoid confusion and misinterpretation the phase should be considered in the family of modulated nematic phases, and thus, named splay nematic phase $N_S$.

### B. The origin of polar order

Berardi et al [11] showed that polar shape with non-equal head-tail interactions can lead to polar nematic phase, and that additional electric dipolar interaction disrupts the polar order. So the obvious question is whether polar order arises primarily because of the shape or because of the electrostatic interaction. As mentioned in Section II, in the case of dipolar interaction, nearest neighbors' positional correlations strongly affect polar ordering. The question is, what positional correlations promote polar order. To answer this, instead of looking at what positional correlations will lead to polar order, we may look at the positional correlation caused by polar order. These can be found in a study of ferrofluids made of spherical nanoparticles with large enough magnetic dipole moment in an external magnetic field [108]. In this study, it has been shown that due to dipolar interaction a pseudocrystalline lamellar hexagonal particle arrangement is formed, in which particles are arranged in hexagonal planes so that the distance between them is about 2 particle diameters, and their magnetic dipole moments lay in the plane of the layer. This arrangement is very similar to what was proposed by N. V. Madhusudana [109] to explain the mechanism for polar order in ferroelectric NLCs. He showed that at high enough density the polar arrangement of elongated constituents with continuously alternating positive and negative charges (which roughly correspond to a line of dipoles with centers separated by twice the dipole size) becomes energetically more favorable than nonpolar arrangement. In the polar state, the neighbors are positionally shifted along the long axis to minimize electrostatic energy (Fig. 17). This electrostatic interaction is very short-ranged so it acts only between the nearest neighbors, and as such, it is obviously different from the one proposed by Born [6]. This mechanism of polar order relies on rather strong positional correlations, which is likely to cause a decrease in molecular mobility. This raises the question of whether such a phase will remain a nematic liquid or will it crystalize. In RM734, a nonpolar crystalline phase is thermodynamically more stable than the polar nematic phase, and in a crystal, a shifted polar nearest neighbors' packing motive is observed although the overall order is nonpolar [18]. On the other hand, the experiments showing that oligomers made of repeating units carrying a dipole form the polar nematic phase [71] support the idea that periodic charge modulation of the constituents promotes polar order.

The question of the origin of polar order remains open. While the (attractive) interactions between molecules are important, one has to bear in mind that the system must remain in a liquid state. This happens when the entropic contributions to the free energy prevail over the attractive interactions. As discussed in Section IV.C., MD studies showed that in materials exhibiting the polar nematic phase, the molecules in the polar phase are more densely packed and at the same time more mobile than in the apolar nematic phase which shows that the polar phase is more favorable from the excluded volume point of view. This demonstrated that the MD studies can be a very useful tool in predicting the existence of the polar nematic phase and supporting the material design. However, simulating the phase transition from apolar to polar nematic phase remains a major challenge for future MD studies. A particular challenge is to incorporate splay deformation in the MD simulation box and check the hypothesis that the splay deformation promotes the polar order and causes the phase transition. The in-depth analysis of the orientational and positional ordering of molecules in MD simulations may also give information on the average potential felt by a molecule and shed a light on how a subtle interplay between polar shape (steric interaction) and weak short-range electrostatic interactions leads to polar order in a liquid.

### C. Development of the theory

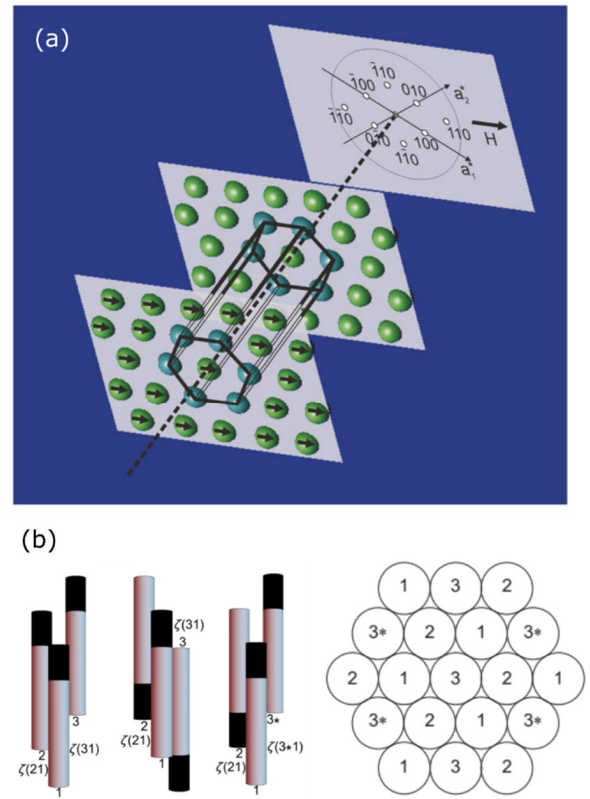

FIG. 17. (a) Schematic representation of a hexagonal arrangement for magnetic spherical particles. ©2003 American Physical Society. Reprinted with permission, Ref. [108] (b) Scheme of polar rod-triplets and their proposed arrangement. © 2021 American Physical Society. Reprinted with permission, Ref. [109].

The structure, field response, and the defects in the nonpolar, achiral or chiral, nematic phase are successfully modeled and predicted by LdG tensor theory or, when the scalar order parameter can be assumed constant, by Oseen-Frank theory. It is a wish and a challenge to establish a similarly successful model to describe and predict the structure, defects, and field response of the ferroelectric nematic phase. This phase is more complicated than apolar nematic because it has lower symmetry and consequently, more terms are allowed in the free energy, among them terms linear in deformation. Which of them are relevant, is a question for future investigations combining experiments and modeling. Additionally, because the electric polarization values are large, the theory needs to include electrostatics. As discussed in Section V and references therein, while in the apolar nematic the surface anchoring can be relatively simply described, the interaction of $N_F$ with the surface is far more complex and sensitive to details, such as pretilt, the position of electrodes and whether they are grounded or floating, the presence of charged impurities, etc. The appropriate description of the surface interaction is a crucial part of the model.

In most of the materials exhibiting the $N_F$ phase, the direction of the dipole and molecular long axis are not parallel and consequently, a polar biaxial phase may form. This interesting, but very complex case was systematically studied by Longa and Trebin [110]. They described the system with coupled biaxial tensor order parameter and polarization, and by using symmetry considerations, they showed that three different biaxial polar phases can exist, one of them chiral. This study demonstrates the complexity with which theoretical description is faced and, on the other hand, it poses a challenge for experimentalists to look for possible other polar nematic phases.

### D. Applications

The final question is the applicability of ferroelectric nematic liquid crystals. This question is of course highly correlated with the development of materials with a suitable temperature range of the $N_F$ phase. However, there is no doubt that these materials will have significant impact on technological implementation of liquid crystals. Although their importance spans from chemistry and biology to physics, in the discussion here, we will focus only on the potential of the physical characteristics in terms of technological applications of the $N_F$ materials. As for the display applications, this question pertains to whether the $N_F$ materials can surpass the performance of the apolar NLCs in terms of speed and voltages needed for optical switching. While the switching of the $N_F$ was shown that it could be of the order of 10-100 μs, using fields $10^5$-$10^6$ V/m [78], the real performance study of the optical response of material on the pixel size, achieving similar intensity (or phase) contrast as in the case of conventional NLCs is still missing. In our opinion, the $N_F$ materials may be also interesting for non-display applications. As discussed in Section IV.A, the $N_F$ materials exhibit nonlinear optic coefficients comparable to those of solid nonlinear optical materials and, provided that in the future, we will learn to control the structures and domains in the $N_F$ phase, they could be used in tunable nonlinear optical components. A remarkable property of $N_F$ materials is their huge temperature-dependent dielectric permittivity. Araoka et al. for example [111], demonstrated that large changes in the dielectric permittivity through the phase transition from $N_S$ to the $N_F$ phase can be achieved also by phototuning and how such behavior can be exploited in electronic applications. It has also been shown that confinement in silanized glass microchannels, either straight or including bend sections, results in continuous $N_F$ order, and thus **P** order, along the channel axis [112]. Such finding by F. Caimi et al. opens the door for the exploration of the use of $N_F$ phases on non-linear fiber optics elements. Also recently, M.T. Máthé et al. have demonstrated thermally induced laminar flow in $N_F$ sessile drops, driven by the unique combination of the tangential orientation of polarization, the pyroelectricity of the fluid and the thermal gradient [113]. Such effect points towards possible exploitation in thermal energy harvesting. Ferroelectric nematic polymers cannot be so easily switched but are due to large NLO coefficients still interesting for application in nonlinear optics and as polar materials with large dielectric permittivity also in electronic devices.

Works collected here constitute only the initial stages of research into applications and exploitable properties of these polar nematic phases. It is expected that this field will rapidly grow with the development of materials.

## VIII. CONCLUSION

The emerging field of ferroelectric nematic liquid crystals has set in motion very exciting times for soft matter research. In this contribution, we recapitulated the different theoretical and experimental relevant works that provide the starting framework for future research development. Research on solid ferroelectric materials is vast and nowadays they play an important role in numerous everyday devices as well as high technological applications, from piezoelectric sensing and dielectric energy storage to electrocaloric solid-state cooling. Although the idea of a ferroelectric liquid dates back to the very beginnings of the discovery of ferroelectricity, hundred years have been needed for their experimental realization. In the liquid ferroelectric state, while retaining mobility typical of a liquid, the molecules spontaneously orient on average in the same direction which results in the appearance of macroscopic electric polarization. Measured spontaneous polarization values ($\sim$ 0.05 C/m$^2$) is comparable to that of some ferroelectric solid materials (e.g. BaTiO$_3$ 0.26 C/m$^2$ [114]). Very large dielectric susceptibility (of the order of 10000), outstanding NLO coefficients ($\sim$5 pm V$^{-1}$) and stronger and faster response to electric fields complete the picture. It is actually the fluid character of the ferroelectric nematic materials, the one that will raise future challenges. Contrary to solid-state ferroelectrics, in an FLC the polarization can vary continuously in space and any restructuring will also be accompanied by material flow. Such key difference speaks for example, for the possibility of tailored polarization structuring through confining surfaces. Finally, the question remains whether the discovery of the ferroelectric nematic phase will lead to further unambiguous realization of long time searched phases as for example the biaxial nematic phase, the splay-twist phase or other polar nematic phases, either predicted, as the 2D splay antiferroelectric phase, or not yet even envisioned. Our success in creating an overarching experimental and theoretical understanding of the already discovered polar nematic phases will play an important role in this subsequent challenge.

Final Note: Research of ferroelectric nematic phases is currently very dynamic and during the editorial processes for the evaluation of this manuscript several new works have been reported and some of them are listed here. New theoretical approaches for the emergence of modulated nematic phases, including $N_S$ phase have been reported by M.P. Rosseto [115] and A. V. Emelyanenko [116]. Several new materials showing ferroelectric phases have been reported, including highly fluorinated and rigid mesogens by Y. Song et al. [117]. Special mention deserves the discovery of new ferroelectric smectic A phases reported by H. Kikuchi et al. [118] and in a preprint by X. Chen et al. [119]. D. Pociecha et al. have shown the inversion of the helical twist in new materials with intrinsically chiral ferroelectric nematic phase at the $N^*$-$N_F^*$ transition [120]. J. Ortega et al. have explored the tenability of multiple bandgaps by low electric fields in chiral ferroelectric nematics [121]. B. Basnet et al. [122] investigate polar in-plane surface interactions as source for domain formation. The behavior of ferroelectric liquid droplets on ferroelectric solid surfaces has been explored by R. Barboza [123].

## ACKNOWLEDGEMENTS

The authors acknowledge the financial support from the Slovenian Research Agency (research core funding No. P1-0192). The authors thank Prof. Satoshi Aya and Prof. Mingjun Huang from South China Advanced Institute for Soft Matter Science and Technology for providing DIO material and Dr. R.J. Mandle from Leeds University for providing RM734 material.

See Supplemental Material at [*URL will be inserted by publisher*] for Movies …..